\documentclass[twocolumn,pre,twoside,preprintnumbers,floatfix]{revtex4} 
\usepackage{graphicx}  
\usepackage{dcolumn} 
\begin{document} 

\title{Structure and thermodynamics of platelet dispersions}
\author{L. Harnau}
\affiliation{Max-Planck-Institut f\"ur Metallforschung,  
Heisenbergstr.\ 3, D-70569 Stuttgart, Germany, 
\\and Institut f\"ur Theoretische und Angewandte Physik, 
Universit\"at Stuttgart, Pfaffenwaldring 57, D-70569 Stuttgart, Germany}

\begin{abstract}
Various properties of fluids consisting of platelike particles differ from the 
corresponding ones of fluids consisting of spherical particles because
interactions between platelets depend on their mutual orientations. One of the 
main issues in this topic is to understand how structural properties of such 
fluids depend on factors such as the shape of the platelets, the size polydispersity, 
the orientational order, and the platelet number density. A statistical mechanics 
approach to the problem is natural and in the last few years there has been a lot 
of work on the study of properties of platelet fluids. In this contribution some 
recent theoretical developments in the field are discussed and experimental 
investigations are described.
\end{abstract}
\maketitle

\section{Introduction}
The symmetry of the shape of platelike particles is reduced compared with 
spherical symmetry so that interactions between them depend not only on 
the separation between their centers but also on their mutual orientations. 
As a result pair correlation functions and phase diagrams of fluids 
consisting of such particles differ from the corresponding ones of fluids 
consisting of spherical particles. In this article various properties 
of platelet fluids are discussed. The structure of the article is as follows.
In Section II some examples of platelet fluids including the experimental 
investigations are presented. The knowledge about the shape of the particles 
and their size polydispersity allows one to calculate intermolecular pair 
correlation functions and the equation of state in the isotropic phase
within the framework of an interaction site integral equation theory as is 
discussed in Sections III and IV. The theoretically obtained static structure 
factor and isothermal compressibility are compared with experimental data. 
Density functional theory is used in Section V to study orientational order in 
spatially homogeneous bulk phases. Moreover, some brief details about properties 
of platelet fluids in contact with substrates are recorded. Some aspects 
of spatially inhomogeneous bulk phases are discussed in Section VI. 
Finally, Section VII draws conclusions.

\section{Shape of particles and polydispersity in size}
In the following we will present examples of systems that illustrate 
the range of interest in platelet dispersions. The shape of the particles 
and the polydispersity in size can be revealed by cryogenic transmission 
electron microscopy and scattering methods.

\subsection{Cryogenic transmission electron microscopy of polyethylene
nanoplatelets}

Polyethylene is a commodity polymer that has become ubiquitous over the past 
several decades because of its low price and good mechanical properties. Hence,
the number of applications of the material is huge and many millions of tons are 
produced worldwide annually. However, polyethylene has hardly played any role in the field 
of nanotechnology. This is due to the problem that polyethylene is produced either by free 
radical polymerization under high pressure and temperature or with metal-organic 
catalysts working exclusively under strictly water-free conditions. Polymer nanoparticles 
and their composites with inorganic compounds, however, are very often produced in 
aqueous systems. Recently, it has been demonstrated that ethylene can be polymerized in aqueous 
systems in a catalytic fashion by Ni(II) complexes. By virtue of this novel synthesis, 
long chains of polyethylene can be generated in a well-controlled environment and at 
ambient temperature. Thus, it could be shown that aqueous polyethylene dispersions can be produced. 
This polymerization has opened the way of well-defined polyethylene nanoplatelets
consisting of the smallest single crystal of polyethylene ever reported \cite{webe:07}. 
Semi-crystalline polyethylene in extremely mild conditions of pressure and temperature 
have been prepared. During this process, nanoplatelets exclusively made of polyethylene are 
formed by the chains as soon as they polymerize. In principle, the crystallization could 
start while the chains are still growing. However, the fast polymerization process and 
the slow nucleation in confined nanoparticles prevent the crystallization from starting 
before the whole particle is formed. It is thus a melt crystallization that takes place 
in each nanoplatelet with an extraordinary degree of supercooling which could not be 
achieved by any other method \cite{chen:07}. 

\begin{figure}[t!]
\begin{center}
\includegraphics[width=9cm, clip=]{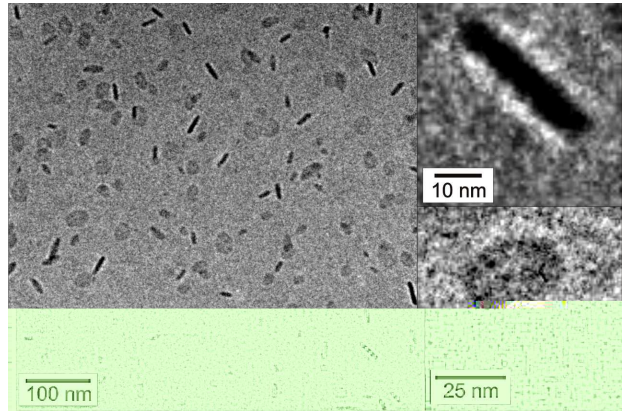}
\caption{Cryogenic transmission electron microscopy micrograph of polyethylene 
nanoplatelets in aqueous dispersion \cite{webe:07}. The grey background is the low-density 
amorphous ice in which the particles are dispersed. The particles are flat platelets, 
appearing as rods when their normal is perpendicular to the electron beam (upper inset) 
and hexagons when their normal is rather parallel to the electron beam (lower inset).}
\label{fig1}
\end{center} 
\end{figure}

Specimens for cryogenic transmission electron microscopy (cryo-TEM) have been prepared by 
vitrification of a thin liquid film of the diluted dispersion supported by a copper 
grid in liquid ethane. Figure \ref{fig1} displays a micrograph of the particles in dilute 
aqueous solution as obtained by cryo-TEM. The dispersion consists of flat platelets. The 
different gray scales for different particles can be easily rationalized by different 
viewing angles: If the normal of a platelet is nearly perpendicular to the electron beam, 
the length of the optical path through the particles is much longer than for parallel 
arrangement. The platelets have a hexagonal shape with rather straight edges. From cryo-TEM 
images a lateral dimension (pseudo-diameter) of \mbox{$25.4\, (\pm \,4.3)$ nm} and a corresponding 
thickness of the platelets of \mbox{$6.3\, (\pm \,0.8)$ nm} have been determined \cite{webe:07}. The 
observed nanoplatelets can be considered as single lamellae of polyethylene. The crystallinity 
of the nanoplatelets has been confirmed by wide-angle x-ray scattering measurements performed 
on the dispersion. The crystalline lamellae visible in Fig.~\ref{fig1} are expected to be 
covered by an amorphous layer as polyethylene is a semicrystalline polymer. However, it 
is not possible to detect this by cryo-TEM. This is due to the fact that the electron density
of amorphous polyethylene is virtually the same as that of the low density amorphous ice. 
Hence, there is not a sufficient contrast between a possible amorphous layer and the surrounding 
medium. However, this problem can be treated by applying contrast variation small-angle x-ray
scattering as an analytical tool as is outlined in the following subsection. Figure \ref{fig1}
also demonstrates that the particles are well-dispersed in the aqueous medium; virtually no
aggregates are found. This is a prerequisite for a meaningful investigation by scattering methods
in solution. 

Cryo-TEM has also been used to analyze the shape and size polydispersity of larger colloidal 
platelets such as nickel hydroxide [Ni(OH)$_2$] platelets of \mbox{180 nm} diameter and  
\mbox{10 nm} thickness \cite{brow:98,brow:99} as well as gibbsite [Al(OH)$_3$] platelets of 
\mbox{200 nm} diameter and  \mbox{15 nm} thickness
\cite{kooi:98,kooi:00a,kooi:00b,kooi:00c,kooi:00d,kooi:01,kooi:01a,beek:03,beek:04a,beek:04,beek:05,wijn:05,wijn:05a,beek:06,mour:06,beek:07}.
However, a quantitative determination of the thickness of these colloidal platelets in 
solution is not straightforward, because in particular thin platelets can hardly ever been 
viewed edgewise on TEM images, while very thick platelets touching the TEM grid with their 
rim have been found. The influence of confining surfaces on the orientation of the 
platelets was also discovered in  suspensions of Ag, Ni, and Fe$_x$Ni$_{1-x}$ nanoplatelets \cite{chen:04,li:04,li:05b,zhan:07}. In order to promote the edgewise orientation of platelets 
on the TEM grid one may use deliberately flocculated samples \cite{kooi:01}. The fact that the 
domains of liquid-crystalline phases of colloidal platelets at higher particle number densities 
extend over rather large distances is also due to the influence of the rigid walls of the 
sample cells.

%
%
%
%

\subsection{Scattering methods}
The theory of static scattering has been presented in various books. Here, we 
review the results necessary for this study. Light scattering, small-angle 
neutron scattering, and small-angle x-ray scattering determine the 
scattering intensity $I({\bf q},\rho)$ as a function of the scattering vector 
${\bf q}$ and the number density of the dissolved particles $\rho$. The absolute 
value of the scattering vector is given by $q=|{\bf q}|=(4\pi n/\lambda)\sin(\theta/2)$ 
in which $n$ is the refractive index of the medium, $\lambda$ is the incident wavelength,
and $\theta$ is the scattering angle. The scattering intensity can be written as 
\begin{eqnarray} \label{eq1}
I({\bf q},\rho)&=&\rho I_0({\bf q}) S({\bf q},\rho)+\rho I_{incoh}\,,
\end{eqnarray}
where  $I_0({\bf q})$ is the 
scattering intensity of a single particle, $S({\bf q},\rho)$ is the structure 
factor related to the mutual interactions of the solute particles, and $I_{incoh}$ 
is the incoherent contribution in small-angle neutron scattering that is mainly due 
to the hydrogen atoms of the particles. The incoherent contribution $I_{incoh}$ must be 
subtracted carefully from experimental data in order to obtain meaningful results on 
the structure and interaction of the dissolved particles. There is no such contribution 
in light scattering and small-angle x-ray scattering. The scattering intensity of a 
single particle may be used to define the form factor $P({\bf q})$ that is normalized 
to unity at $q=0$ according to
\begin{eqnarray} \label{eq2}
I_0({\bf q})&=&V^2_p (\Delta {\tilde\rho})^2 P({\bf q})\,,
\end{eqnarray}
where $V_p$ is the volume of a dissolved particle and $\Delta {\tilde \rho}$ is the 
contrast of the solute resulting from the difference of the average scattering length 
density of the dissolved particles and the scattering density of the solvent. In the 
case of light scattering and small-angle x-ray scattering, the contrast is fixed by the 
solvent, the partial specific volume of the solute, and its chemical composition. In the 
case of small-angle neutron scattering, $\Delta {\tilde \rho}$ is an available model 
parameter that can be varied by mixing protonated and deuterated solvent. The form 
factor of randomly oriented monodisperse circular platelets of radius $R$ and thickness 
$L$ is given by 
\begin{eqnarray} \label{eq3}
P(q)&=&4\int\limits_0^1 dx\,\frac{J^2_1(qR\sqrt{1-x^2})}{(qR\sqrt{1-x^2})^2}
\frac{\sin^2(qLx/2)}{(qLx/2)^2}\,,
\end{eqnarray}
where $J_1(x)$ denotes the cylindrical Bessel function of first-order. The effect 
of size polydispersity of the platelets is taken into account by an appropriate average 
according to
\begin{eqnarray} \label{eq4}
\left\langle I_0({\bf q}) \right\rangle &=&\int\limits_0^\infty dR\,\int\limits_0^\infty dL\,
I_0({\bf q})G(R,L)\,,
\end{eqnarray}
where $G(R,L)$ is a distribution function characterizing the degree of polydispersity.

\subsubsection{Form factor of stilbenoid dendrimers}
\begin{figure}[t!]
\begin{center}
\includegraphics[width=8cm, clip=]{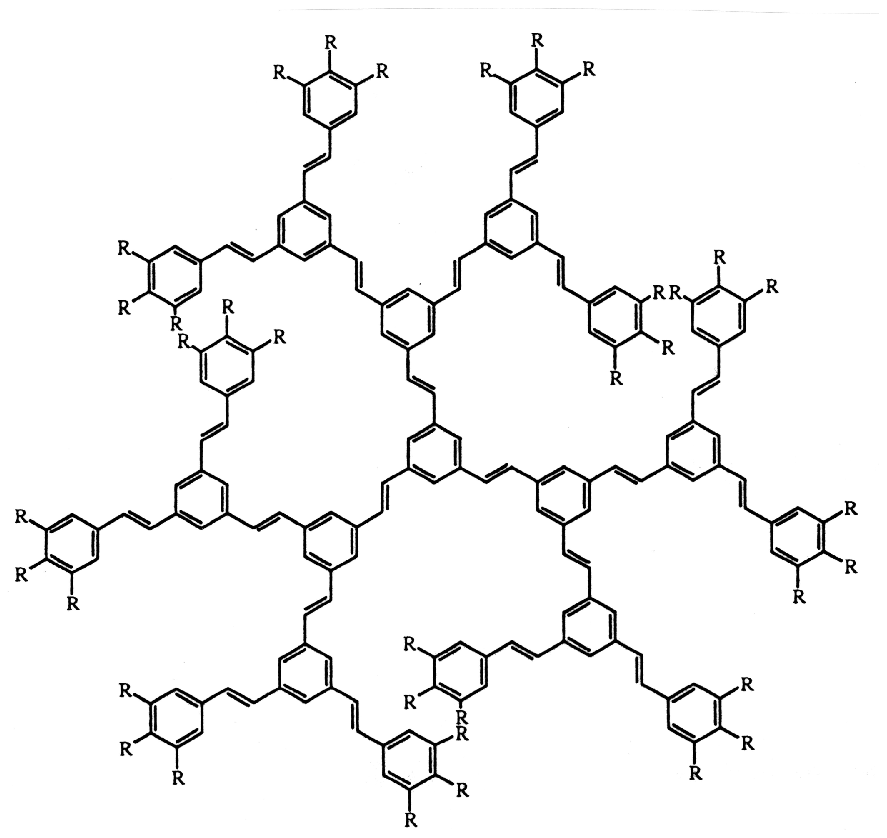}
\caption{Chemical structure of a stilbenoid dendrimer of the third generation
\mbox{(R $ = OC_6H_{13}$).}}
\label{fig2}
\end{center} 
\end{figure}

Dendrimers are synthetic macromolecules with defined architecture that are synthesized 
by iterative controlled reaction steps. Starting from a focal unit, subsequent generations 
are connected to this initial core which results in a treelike structure. Dendrimers 
composed of flexible units adopt a so-called dense core structure characterized by 
an average segmental density that has its maximum at the center of the 
molecule. This is easily derived from the fact that flexible dendrimers can assume a 
great number of conformations in which the terminal groups can fold back. Hence these 
flexible dendrimers do not exhibit a well-defined surface given by the terminal groups 
of the last generation. The average density profile thus derived can be used to calculate 
the interaction of flexible dendrimers in solution. These structures are well understood 
by now \cite{ball:04}. Much less attention has been paid to dendrimers consisting of rigid 
units \cite{meie:98,muel:99,meie:00,wind:02,meie:03,rose:04,rose:05,carb:06,schu:07}. 
Figure \ref{fig2} shows the chemical structure of a rigid stilbenoid dendrimer of 
the third generation which is solely composed of stiff units. Starting from a central 
phenyl group, all subsequent generations are built up by trans-stilben units, only the 
terminal groups are substituted by hexyloxy groups in order to ensure better solubility 
in common solvents such as toluene. Full conjugation in trans-stilbene can be achieved in 
the completely planar conformation. However, the potential energy for a slight torsion around 
the single bonds is low in the ground state of trans-stilbene. Molecular modeling without 
taking into account the solvent has demonstrated that a stilbenoid dendrimer of the third 
generation exhibits a relatively compact platelike structure \cite{meie:00}.

%
%
%
%

The spatial structure of the stilbenoid dendrimers in dilute solution has been 
investigated by small-angle x-ray scattering and small-angle neutron scattering 
\cite{rose:06}. Both methods give different information: Small-angle neutron 
scattering data taken at high contrast between the solute and the solvent can be 
used to explore the shape of the entire molecule. Small-angle x-ray scattering, 
however, is mainly sensitive to the stilbenoid scaffold of the molecule and hardly 
detects the hydrocarbon chains attached to its periphery due to the fact that 
these hexyl chains are virtually matched by the solvent toluene.

\begin{figure}[t!]
\begin{center}
\includegraphics[width=8cm, clip=]{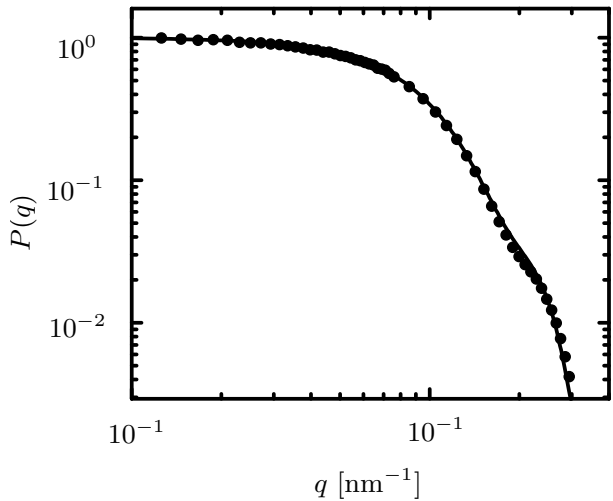}
\caption{The form factor $P(q)$ of monodisperse stilbenoid dendrimers of the third 
generation \cite{rose:06} as obtained by small-angle neutron scattering 
(circles) and calculated according to Eq.~(\ref{eq3}).}
\label{fig3}
\end{center} 
\end{figure}

The form factor $P(q)$ of stilbenoid dendrimers of the third generation as obtained 
by small-angle neutron scattering is shown in Fig.~\ref{fig3}. The form factor agrees 
with the one calculated numerically for a monodisperse circular platelet with a radius 
\mbox{$R=2.4$ nm} and thickness \mbox{$L=1.8$ nm} according to Eq.~(\ref{eq3}). The 
good agreement between the experimental and theoretical results demonstrates that a 
full understanding of the spatial structure of stilbenoid dendrimers of the third 
generation has been achieved \cite{rose:06}. Dendrimers consisting of rigid units 
exhibit a rather well defined structure in solution and may serve as model systems 
for interacting monodisperse particles in statistical physics \cite{harn:07} 
as is discussed in Section \ref{ISO}. 

%
%
%
%

\subsubsection{Form factor of laponite platelets}
Laponite consists of platelike clay particles with a radius \mbox{$R \approx 11.5$ nm}
and thickness \mbox{$L \approx 0.9$ nm}. Each platelet carries a few hundred 
elementary charges. For a long time, aqueous suspensions of laponite have been 
extensively investigated as a model system for platelike colloids
\cite{Ram:86,Ram:90,Ram:93,chan:94,Mou:95,Kro:96,triz:97a,triz:97b,hsu:97,dijk:97,Kro:98,Mou:98a,Mou:98b,delv:99,Bon:99a,Bon:99b,rowa:00,kutt:00,Kna:00,Bon:01,harn:01,nico:01a,nico:01b,Bon:02,triz:02,Mar:02,Bell:03,Tan:04,agra:04,Mon:04,ruzi:04a,band:04,odri:04,Tan:05,Mon:05,ruzi:06,scho:06,delv:06,mart:06,moss:07,cumm:07,iann:07,ruzi:07,faro:07}. 
A theoretical description of such a 
well-characterized model clay appeared feasible. Early attempts focused on 
the structure of the electric double layer around a single platelet 
\cite{chan:94,triz:97a,triz:97b}, or around two parallel platelets \cite{delv:99}. 
Much of the recent theoretical work was initiated by considering laponite platelets 
and their associated electric double layers as nonintersecting platelets carrying
a constant electrostatic quadrupole moment \cite{dijk:97}. The spatial arrangement 
of these objects has been investigated by Monte Carlo simulations. The results of 
these studies point to a reversible sol to gel transition at platelet number density 
$\rho > 0.06\, R^{-3}$, where the structure of the gel phase is strongly reminiscent of 
a "house-of-cards".

A screened electrostatic potential between two arbitrary oriented, charged 
platelets has been worked out within the framework of a linearized 
Poisson-Boltzmann theory \cite{hsu:97,rowa:00,triz:02,agra:04}. At a fixed 
center-to-center distance, this repulsive electrostatic potential
is maximized for coplanar platelets, which corresponds to the maximum overlap 
of electric double layers, and minimized when the platelets are coaxial and 
parallel. An intermediate potential is found for T-shaped perpendicular platelets. 
At present only preliminary results of Monte Carlo simulations for platelets 
interacting through such an anisotropic potential are available, indicating 
that the platelet fluid is in the isotropic phase for platelet number density 
$\rho < 0.1\, R^{-3}$ \cite{triz:02}. Later an interaction site model of laponite was 
put forward, and investigated by molecular-dynamics simulations \cite{kutt:00} and 
integral equations \cite{harn:01}. In this model each platelet carries discrete 
charged sites rigidly arranged on an array inscribed in the platelet. Sites on different 
platelets interact via a repulsive screened Coulomb interaction. Upon varying the 
platelet number density and screening length, a sol, gel, and crystal phase have 
been identified. Additional rim charges of opposite sign have been found to lead to 
T-shaped pair configurations and clustering of the platelets.

\begin{figure}[t!]
\begin{center}
\includegraphics[width=8.4cm, clip=]{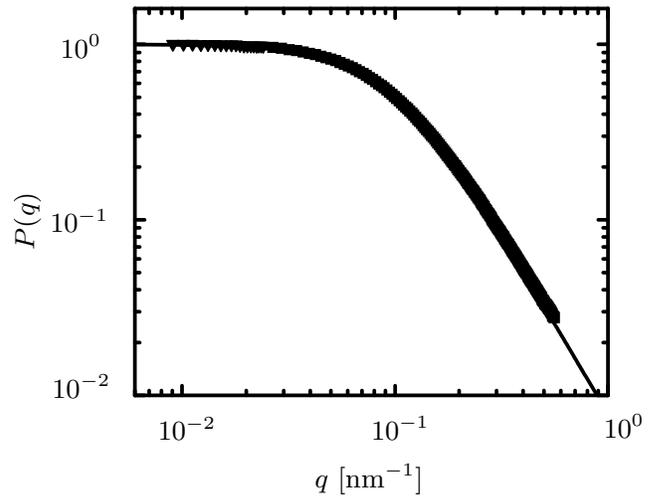}
\caption{The form factor $P(q)$ of polydisperse laponite platelets \cite{li:05} as 
obtained by light scattering (triangles) and small-angle x-ray scattering (squares).
The solid line represents the theoretical results for polydisperse circular 
platelets according to Eqs.~(\ref{eq2}) - (\ref{eq4}).}
\label{fig4}
\end{center} 
\end{figure}

The theoretical studies on gelation and structure formation in laponite suspensions 
have assumed a system of monodisperse platelets, i.e., thin platelets with a radius 
$R$ and a thickness $L$. However, it has become possible to visualize directly 
the laponite platelets by various techniques such as atomic force microscopy or electron 
microscopy. From these micrographs it is obvious that laponite exhibits a considerable 
size polydispersity \cite{Bih:01,Bal:03,Her:04,leac:05}. The degree of polydispersity 
has been quantified in a comprehensive study of laponite suspensions in dilute
solution using light scattering and small-angle x-ray scattering \cite{li:05}. 
Figure \ref{fig4} displays the measured and 
calculated form factor $P(q)$ of laponite platelets. From a comparison of Fig.~\ref{fig3} 
and Fig.~\ref{fig4} it is obvious that size polydispersity must be included in the case of 
laponite platelets because the form factor of monodisperse platelets given by 
Eq.~(\ref{eq3}) has a wavy shape which is smoothed out by size polydispersity in the case 
of the laponite platelets. The solid line in Fig.~\ref{fig4} shows the resulting fit 
obtained for a Schulz-Zimm-distribution with a polydispersity expressed through 
$R_w/R_n = 1.6$. The weight-average radius $R_w$ is given by \mbox{$11.5$ nm}, the 
thickness is found to be \mbox{L=$0.9$ nm}, and the number-average radius is denoted by 
$R_n$. It is worthwhile to reiterate that laponite platelets exhibit an appreciable 
polydispersity that can be deduced from this analysis. Obviously, this must be taken 
into account when modeling the interaction of the particles at finite volume fraction.

%
%
%
%

\section{Intermolecular pair correlations in the isotropic phase}
Using the form factor as input into generalized Ornstein-Zernike equations of an 
interaction site integral equation theory, one can calculate structural properties of 
interacting platelets. In this section we will focus on results obtained from 
intermolecular pair correlation functions in the isotropic phase. Calculated structure 
factors will be compared with experimental data. 

\subsection{RISM and PRISM theory}
The systems under investigation are dispersions or solutions, but in view of the 
mesoscopic scale of the particles, the solvent will be considered as a structureless 
continuum. Spatial pair correlations of an isotropic fluid of identical particles, 
each carrying $n$ distinct interaction sites, are characterized by a set of 
intermolecular site-site total correlation functions $h_{ij}(r,\rho)$, where the 
indices $i$ and $j$ run over sites on each of two particles, and $\rho$ is the 
particle number density. These functions are related to a set of intermolecular 
site-site direct correlation functions  $c_{ij}(r,\rho)$ by the generalized 
Ornstein-Zernike relations of the ''reference interaction site model'' (RISM), 
which in Fourier space read \cite{chan:72,chan:82}
\begin{eqnarray} 
h_{ij}(q,\rho)&=&\sum\limits_{m, o =1}^n\omega_{im}(q)c_{mo}(q,\rho)
\left(\omega_{oj}(q)+\rho h_{oj}(q,\rho)\right)\,,\nonumber
\\&&        \label{eq5}
\end{eqnarray}
where the $\omega_{ij}(q)$ are the Fourier transforms of the intramolecular 
correlation functions for rigid particles. For flexible particles the intramolecular 
correlation functions depend on the particle number density and follow from a statistical 
average over particle configurations. In the case of rigid particles such as the 
platelets considered here, the $\omega_{ij}(q)$ are independent of the particle number 
density because the particles are not deformed due to intermolecular interactions 
for typical concentrations in the fluid state. The set of generalized Ornstein-Zernike 
equations must be supplemented by a set of closure relations. If the interaction sites 
are simply the centers of exclusion spheres of diameter $d$, to account for steric effects, 
a convenient closure is the Percus-Yevick approximation \cite{chan:72,hans:86}
\begin{eqnarray} \label{eq6}
h_{ij}(r,\rho)=-1\,,\,\,r\le d\,,\hspace*{0.5cm}
c_{ij}(r,\rho)=0\,,\,\,r> d\,.
\end{eqnarray}
The experimentally accessible structure factor $S(q,\rho)$ is defined as
\begin{eqnarray} \label{eq7}
S(q,\rho)=1+\rho\frac{h(q,\rho)}{P(q)}\,,
\end{eqnarray}
where 
\begin{eqnarray} \label{eq8}
h(q,\rho)=\frac{1}{n^2}\sum\limits_{m, o =1}^nh_{mo}(q,\rho)
\end{eqnarray}
is the particle-averaged total correlation function. The particle-averaged 
intramolecular correlation function
\begin{eqnarray} \label{eq9}
P(q)=\frac{1}{n^2}\sum\limits_{m, o =1}^n\omega_{mo}(q)
\end{eqnarray}
characterizes the geometry of the distribution of the sites, and hence the geometric 
shape of the particles. While the particle-averaged intramolecular correlation function 
accounts for the interference of radiation scattered from different parts of the same 
particle in a scattering experiment, the local order in the fluid is characterized by 
$h(q,\rho)$ or $S(q,\rho)$. 

The RISM has been proved to be a successful theory of the pair structure of many molecular 
fluids (for a review see Ref.~\cite{mons:90}). In the case of macromolecular and colloidal systems, with very large numbers of interaction sites, the number of coupled RISM equations
becomes intractable, and a considerable simplification follows from the assumption that the
direct correlation functions $c_{ij}(q,\rho)$ are independent of the indices $i$ and $j$. 
This leads to the ''polymer reference interaction site model'' (PRISM) theory first applied 
by Schweizer and Curro to long flexible polymers \cite{schw:87}. PRISM neglects end effects 
in that case. The resulting single generalized Ornstein-Zernike equation of the PRISM reads 
\begin{eqnarray} \label{eq10}
h(q,\rho)=P(q)c(q,\rho)P(q)+\rho c(q,\rho)h(q,\rho)P(q)\,,
\end{eqnarray}
where $c(q,\rho)=\sum_{m,o=1}^nc_{mo}(q,\rho)$. Taking into account size polydispersity 
leads to coupled generalized Ornstein-Zernike equations of the multicomponent system 
consisting of $N$ species:
\begin{eqnarray}
h_{\alpha \gamma}(q,\{\rho_\alpha\})&=&P_\alpha(q)
c_{\alpha \gamma}(q,\{\rho_\alpha\})P_\gamma(q)\nonumber
\\&+&P_\alpha(q)\sum\limits_{\delta=1}^N
c_{\alpha \delta}(q,\{\rho_\alpha\})
\rho_\delta h_{\delta \gamma}(q,\{\rho_\alpha\})\,.\nonumber
\\&&  \label{eq11}
\end{eqnarray}
This set of $N(N+1)/2$ independent Ornstein-Zernike equations must be supplemented by 
as many closure relations between each pair of total and direct correlation functions.
The PRISM integral equation theory has been successfully applied to various systems, such 
as rodlike viruses \cite{yeth:96,yeth:97,yeth:98,harn:00}, polymers \cite{schw:97,harn:01a}, 
mixtures of spherical colloids and semiflexible polymers \cite{harn:02}, bottlebrush 
polymers \cite{boli:07}, and polyelectrolyte brushes \cite{henz:08}. Moreover, it has been
demonstrated that the simpler PRISM theory yields results in good agreement with the more 
elaborate RISM calculations for lamellar colloids \cite{cost:05}.

\subsection{Structure factor of stilbenoid dendrimers}
\begin{figure}[t!]
\begin{center}
\includegraphics[width=8cm, clip=]{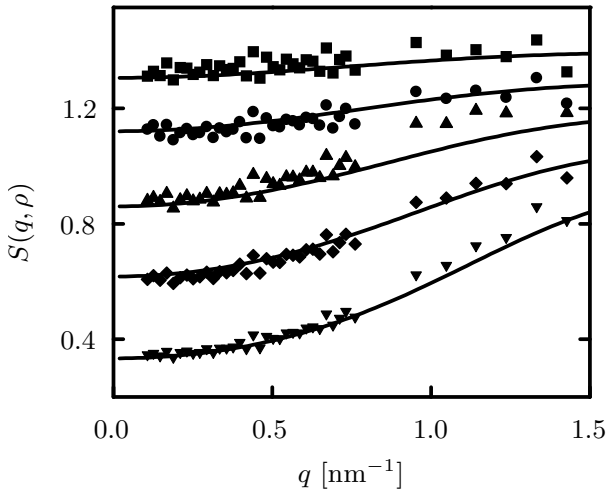}
\caption{Experimentally determined structure factors $S(q,\rho)$ for monodisperse
stilbenoid dendrimers of the third generation (see Fig.~\ref{fig2}) together with 
the results of the theoretical predictions of the PRISM integral equation theory according 
to Eqs.~(\ref{eq3}),  (\ref{eq7}), and (\ref{eq10}) \cite{harn:07}. The volume fraction 
of the dendrimers increases from top to bottom: 
$\phi=0.009, 0.019, 0.039, 0.061, 0.1$. For reasons of clarity, the upper data sets 
and lines have been shifted up by $0.1$, $0.2$, $0.3$, $0.4$, respectively.}
\label{fig5}
\end{center} 
\end{figure}

In Fig.~\ref{fig5} the experimental structure factors $S(q,\rho)$ for monodisperse
stilbenoid dendrimers of the third generation (see Fig.~\ref{fig2}) are compared to 
the results of the integral equation theory for the PRISM \cite{harn:07}. The 
calculated form factor $P(q)$ (see the solid line in Fig.~\ref{fig3}) has been
used as input into the generalized Ornstein-Zernike equation [Eq.~(\ref{eq10})]. 
The volume fraction is given by $\phi=V_p\rho$, where $V_p$ is the volume 
of an individual particle. The generalized Ornstein-Zernike equation has been
solved numerically together with the Percus-Yevick closure. From Fig.~\ref{fig5} 
it is apparent that the PRISM integral equation theory is rather accurate. The 
magnitude and the scattering vector range of the suppression of $S(q,\rho)$, i.e., 
the deviations from the value 1 at small scattering vectors, are characteristic 
for the size and the shape of the dendrimers as well as the volume fraction. 
On the basis of the experience with both PRISM and RISM it is expected that 
the results of the integral equation theory for the RISM would lead to very 
similar results, provided the same form factor is used \cite{cost:05}.
%
%
%
%

\subsection{Scattering intensity of polyethylene nanoparticles}
The above mentioned dispersions consisting of polydisperse polyethylene nanoplatelets 
(see Fig.~\ref{fig1}) contained in addition just enough of the surfactant sodium 
dodecyl sulfate to stabilize the particles against coagulation \cite{webe:07}. The 
surface tension of \mbox{65 mN/m} of the aqueous dispersion demonstrated that virtually 
all the surfactant was adsorbed onto the particles leading to an effective charge of the 
platelets which has to be taken into account in the PRISM integral equation theory. One 
may consider a multicomponent system involving $N$ species of charged platelets with 
number densities $\rho_\alpha$, where $1\le \alpha \le N$. Each platelet of species 
$\alpha$ contains $n_\alpha$ equivalent interaction sites. The interaction potential 
between sites on particles of species $\alpha$ and $\gamma$, carrying the charges 
$z_\alpha e$ and $z_\alpha e$, is of the generic form: 
\begin{equation}  \label{eq12}
U_{\alpha \gamma}(r)=\frac{\displaystyle z_\alpha z_\gamma e^2}{\displaystyle\varepsilon r} 
\exp(-\kappa_Dr)\,, 
\end{equation} 
where $\lambda_D=\kappa_D^{-1}$ is the usual Debye screening length. The system is an 
aqueous dispersion, but in view of the mesoscopic scale of the 
platelets, the solvent is modelled as a structureless dielectric continuum providing a 
macroscopic permittivity $\varepsilon$. Any microscopic counterions or ions from added 
electrolyte will be considered at the linear response (or Debye-H\"uckel) level, i.e., 
they will screen the electrostatic potential due to the interaction sites on the charged 
platelets on a scale given by the Debye screening length. The underlying assumption 
entails that the charge distribution on the mesoscopic particles does not contribute to 
screening. 

\begin{figure}[t!]
\begin{center}
\includegraphics[width=8cm, clip=]{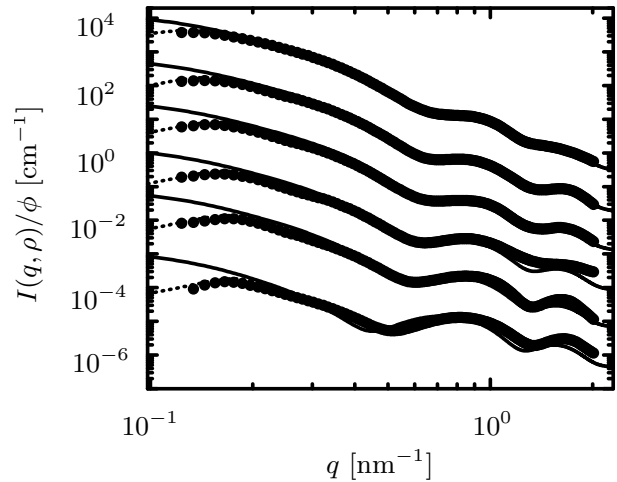}
\caption{Measured scattering intensity of polyethylene nanoplatelets as a function of 
the magnitude of the scattering vector $q$ (symbols) \cite{webe:07}. All intensities have been 
normalized to the volume fraction $\phi$ of the platelets in the dispersion. The volume fraction 
of the added sucrose increases from bottom to top (0, 6.2, 10.3, 18.0, 25.4, 37.8 vol.\%) while 
the volume fraction $\phi$ of the nanoplatelets decreases from bottom to top 
(1.73, 1.56, 1.44, 1.23, 1.02, 0.68 vol.\%). The five lowermost intensities are shifted down 
by a factor of $10$, $10^2$, $10^3$, $10^4$, $10^5$, respectively. The solid lines represent 
the result of the modeling of the small-angle x-ray scattering data assuming a dispersion of noninteracting 
polydisperse platelets. The short dashed lines ($q < 0.25$ nm$^{-1}$) represent the scattering 
intensity calculated for a dispersion of interacting platelets as obtained from the PRISM integral 
equation theory [Eqs.~(\ref{eq11}) - (\ref{eq14})]. The differences between the dotted and 
solid lines reflect the intermolecular interaction between the nanoplatelets.}
\label{fig6}
\end{center} 
\end{figure}

The Laria-Wu-Chandler closure relation \cite{lari:91} has been used in order to 
supplement the Ornstein-Zernike equations [Eq.~(\ref{eq11})] of the multicomponent system:
\begin{eqnarray}
h_{\alpha \gamma}(r,\{\rho_\alpha\})&=&\ln[h_{\alpha \gamma}(r,\{\rho_\alpha\})+1]\nonumber
\\&+&P_\alpha \star[c_{\alpha \gamma}+(k_BT)^{-1}n_\alpha n_\gamma
U_{\alpha \gamma}]\star P_\gamma(r)\,,
\nonumber
\\&&\label{eq13}
\end{eqnarray} 
where the asterisk $\star$ denotes a convolution product. In addition, steric 
effects have been taken into account using the Percus-Yevick approximation. 
The cryo-TEM micrograph displayed in Fig.~\ref{fig1} shows that the particles are hexagonal 
platelets with a pseudo-radius of 12.5 nm and a thickness of 6.3 nm. As argued above, 
however, the cryo-TEM micrograph shows only the crystalline lamella. Since polyethylene is 
a semicrystalline polymer, an amorphous layer in which the chains fold back must exist on 
both sides of the lamella. Hence, there must be a thin amorphous layer on both sides of the 
particles that needs to be taken into account. The form factor of such a ''hamburger'' is 
given by
\begin{eqnarray} \label{eq14}
P(q)&=&4\int\limits_0^1 dx\,\frac{J^2_1(qR\sqrt{1-x^2})}{(qR\sqrt{1-x^2})^2}
\left(L\frac{\sin(qLx/2)}{(qLx/2)}\Delta {\tilde\rho}_1\right.\nonumber
\\&+&\left.L_c\frac{\sin(qL_cx/2)}{(qL_cx/2)}
(\Delta {\tilde\rho}_c-\Delta {\tilde\rho}_1)\right)^2\nonumber
\\&\times&\left(L\Delta {\tilde\rho}_1+
L_c(\Delta {\tilde\rho}_c-\Delta {\tilde\rho}_1)\right)^{-2}\,.
\end{eqnarray}
Here $R$ denotes the radius of the platelet, $L$ is the overall thickness of the particles, 
while $L_c$ denotes the thickness of the crystalline lamella. $\Delta {\tilde\rho}_1$
and $\Delta {\tilde\rho}_c$ are the contrasts of the outer two platelets and of the inner platelet, 
respectively. The polydispersity of the particles estimated from the cryo-TEM micrographs has been 
taken into account in the fitting using a Gaussian distribution in radius and thickness. The closed 
set of equations (\ref{eq11}) - (\ref{eq14}) has been solved numerically by a standard iterative 
procedure in order to obtain the scattering intensity \cite{webe:07}. The solid lines in 
Fig.~\ref{fig6} display the form factors. The small-angle x-ray scattering intensities of the 
polyethylene nanoplatelets have been measured at six different 
contrasts starting from a stock solution of \mbox{$\phi= 0.017$} of the nanoplatelets dispersed in 
pure water. The different contrasts have been adjusted by adding different amounts of sucrose. 
The volume fractions of added sucrose increased from 0.0 (lowest contrast) to 0.378 (highest contrast) 
while the corresponding volume fraction of the nanoparticles decreased from \mbox{$\phi= 0.017$}
to \mbox{$\phi= 0.007$}. For the sake of clarity the scattering intensities related to different 
contrast have been shifted vertically in Fig.~\ref{fig6}. The figure demonstrates that varying 
the contrast leads to marked differences in the scattering intensities. In particular, the maxima 
of the scattering intensities are shifted in a characteristic manner when changing the contrast. 
The overall dimensions following from the theoretical description are the weight-average radius 
\mbox{$R_w = 12$ nm} and the weight-average thickness \mbox{$L_w = 9$ nm}. The thickness of the 
crystalline layer is \mbox{$L_c = 6.3$ nm}. From the particle volume, the polyethylene density and 
the chain molecular weight, one can estimate that each particle is made up of ca. 14 chains. 
Moreover, from the entire series of small-angle x-ray scattering-intensities the contrast results to 
\mbox{$\Delta {\tilde\rho}_1 = 309$ electrons/nm$^3$} for the amorphous part and 
\mbox{$\Delta {\tilde\rho}_1 = 333$ electrons/nm$^3$} for the crystalline part. The electron 
densities of the crystalline and amorphous parts of the particles can be used to understand the 
cryo-TEM micrograph in detail. The value for $\Delta {\tilde\rho}_1$ is close to the electron 
density of low density amorphous-ice. Therefore one can conclude that the amorphous parts are
virtually matched in the cryo-TEM.

For small magnitudes of the scattering vector $q$ the calculated scattering intensities for 
noninteracting particles (solid lines) on the one hand, and the integral equation results for 
interacting particles (dotted lines) as well as the experimental data (symbols) on the other 
hand deviate due to strong repulsive interactions between the particles. The observed decrease of 
the scattering intensity at small scattering vectors is considerably more pronounced than for fluids
consisting of hard platelets such as the rigid dendrimers discussed above. From this result 
one can conclude that the colloidal stability is achieved by electrostatic repulsion between the 
platelets brought about by the adsorbed surfactant. It has been estimated that each polyethylene nanoplatelet 
is surrounded by a layer consisting of a few hundred surfactant molecules \cite{webe:07}.

%
%
%
%

\subsection{Structure factor of laponite}
\begin{figure}[t!]
\begin{center}
\includegraphics[width=8cm, clip=]{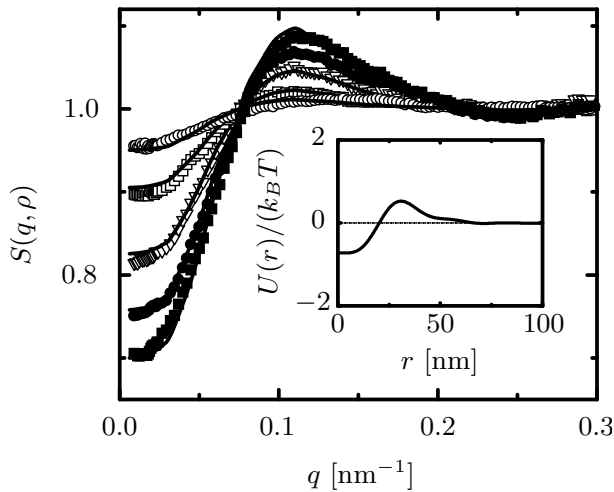}
\caption{The structure factor $S(q,\rho)$ of laponite platelets for various 
volume fractions (open circles: 0.02 $\%$; open squares: 0.04 $\%$; 
open triangles: 0.08 $\%$; solid circles: 0.12 $\%$; solid squares: 0.16 $\%$). 
The solid lines represent the results of the integral equation theory 
[Eqs.~(\ref{eq15}) - (\ref{eq17})] using the effective potential $U(r)$ shown in 
the inset \cite{li:05}.}
\label{fig7}
\end{center} 
\end{figure}

While the preceding subsections have demonstrated that the PRISM theory can 
describe pair correlations of both uncharged and uniformly charged platelike 
particles in solution, disagreement between experimental and theoretical results 
has been found in the case of laponite platelets if  a purely repulsive screened
Coulomb potential is used in the theoretical calculations \cite{li:05}. It has 
been suggested \cite{ruzi:04} that at low concentrations the effective interaction 
between laponite platelets is characterized by a competition of long-range 
Coulomb repulsion and short-range van der Waals attraction similar to colloidal 
systems and protein solutions \cite{piaz:00,pell:03,scio:04}. Moreover, the charge 
density is higher on the face than on the rim of a platelet, which leads to a 
modification of the interaction potential between sites on particles within the
computationally demanding multicomponent PRISM theory. Therefore, a mesoscopic coarse
graining, whereby particles act via an effective potential $U(r)$ has been proposed 
\cite{li:05}. To this end the Ornstein-Zernike equation 
\begin{equation} \label{eq15}
h(q,\rho)=c(q,\rho)+\rho c(q)h(q,\rho)
\end{equation} 
has been solved together with the hypernetted-chain closure relation 
\begin{equation} \label{eq16} 
h(r,\rho)=\ln[h(r,\rho)+1]+c(r,\rho)+(k_BT)^{-1}U(r)\,.
\end{equation} 
These equations follow from Eqs.~(\ref{eq11}) and (\ref{eq13}) for $N=1$ and $P_\alpha(q)=1$. The 
calculated structure factor 
\begin{equation} \label{eq17} 
S(q,\rho)=1+\rho h(q,\rho) 
\end{equation} 
is compared in Fig.~\ref{fig7} with experimental data \cite{li:05}. For all volume fractions, 
good agreement with the experimental results has been achieved with a density-independent 
effective potential $U(r)$ that is attractive for small distances but repulsive for larger 
distances (see the inset in Fig.~\ref{fig7}). The effective potential fulfills the stability 
condition $\int_0^\infty dr\,r^2 U(r)>0$ discussed by Ruelle \cite{ruel:69} and provides 
a representation of the underlying many-body interactions in the system. In general,
to any $S(q,\rho)$ there corresponds a unique effective 
center of mass potential $U_{eff}(r,\rho)$, capable of reproducing $S(q,\rho)$, irrespective 
of the underlying many-body interactions in the system. This effective potential is rather 
independent of the particle number density for the laponite suspensions, i.e., 
$U_{eff}(r,\rho)\approx U(r)$.

%
%
%
%

\begin{figure}[t!]
\begin{center}
\includegraphics[width=8cm, clip=]{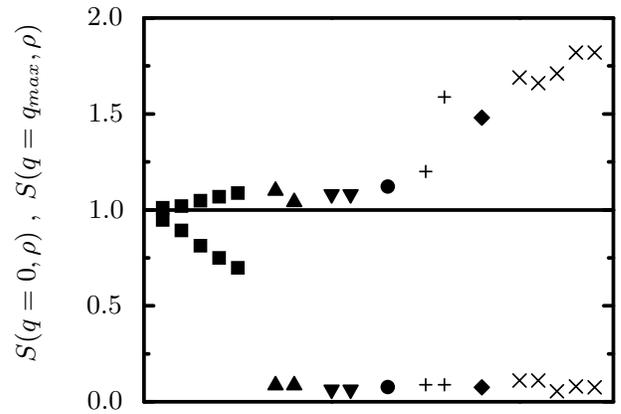}
\caption{The structure factor at zero scattering vector \mbox{$S(q=0,\rho)$} 
(lower symbols) together with the structure factor at the main peak $S(q=q_{max},\rho)$ 
(upper symbols) for various suspensions: laponite at five volume fractions: 
\mbox{0.02 $\%$}, \mbox{0.04 $\%$}, \mbox{0.08 $\%$}, \mbox{0.12 $\%$}, 
\mbox{0.16 $\%$} from left to right (squares \cite{li:05}); 
polystyrenesulfonate of length $L=6.8$ nm at two concentrations: 0.1 mol/L and 
0.2 mol/L from left to right (up triangles \cite{kass:97}); 
polystyrenesulfonate of length $L=40$ nm at two concentrations: 0.1 mol/L and 
0.2 mol/L from left to right (down triangles \cite{kass:97}); 
DNA of length $L=57$ nm at $0.05$ mol/L (circles \cite{kass:97}); 
proteoglycan at two different salt concentrations: 0.05 mM and 0 mM 
from left to right (plus-symbols \cite{norw:96}); 
DNA of length $L=380$ nm at $0.05$ mol/L (diamonds \cite{kass:97}); 
tobacco mosaic virus at five concentrations: 
0.11 mg/mL, 0.27 mg/mL, 0.43 mg/mL, 1.05 mg/mL, 2.07 mg/mL 
from left to right (crosses \protect\cite{maie:88}).}
\label{fig8}
\end{center} 
\end{figure}

Since structure factors and pair correlation functions of polyelectrolyte solutions 
calculated within the PRISM framework and using a purely repulsive screened
Coulomb potential are in good agreement with experimental data and computer simulations, 
it is worthwhile to compare scattering data for polyelectrolytes with the data obtained 
for laponite. Figure \ref{fig8} displays $S(q=0,\rho)$ together with 
$S(q=q_{max},\rho)$ for various polyelectrolytes and laponite \cite{li:05}, where $q_{max}$ 
is the absolute value of the scattering vector at the main peak. The structure factors of the 
polyelectrolytes exhibit a main peak $S(q=q_{max},\rho)>1$ and a small value 
$S(q=0,\rho)<0.11$ for small scattering vectors, while the structure factors of laponite 
are characterized by a main peak and a rather large value $0.7<S(q=0,\rho)<0.95$ for 
small scattering vectors. The enhancement of the small angle 
($q\to 0$) value of $S(q,\rho)$ of laponite as compared to the polyelectrolytes signals 
increased density fluctuations. Since the polyelectrolyte solutions remain liquid like even
at higher concentrations opposite to the laponite suspensions, the observed qualitative 
different behavior of the structure factors of laponite may be considered as indicative of 
the sol to gel transition at higher concentrations. In addition the values of 
$S(q=0,\rho)$ for the liquid like laponite suspensions are smaller than 1 and decrease 
with increasing concentration in contrast to strong small $q$ upturns which have been 
observed experimentally for clay suspensions in the gel phase, 
polyelectrolyte gels and mixtures, and  low ionic strength polyelectrolyte solutions. 
These strong upturns signal strong concentration fluctuations indicative of aggregation, 
or spinodal instability reminiscent of the behavior observed in computer 
simulations and PRSIM integral equation theory of mixtures 
of oppositely charged  particles \cite{harn:02}. 

%
%
%
%

\section{Equation of state and isothermal compressibility in the isotropic phase} 
\label{ISO}

The structure factor provides a direct link with thermodynamics via the compressibility 
equation \cite{hans:86}
\begin{eqnarray} \label{eq18}
\lim\limits_{q\to 0}S(q,\rho)
=\rho k_BT\kappa_T(\rho)\,,
\end{eqnarray}
where $\kappa_T(\rho)$ is the isothermal compressibility. The osmotic pressure 
$P(\rho)$ (equation of state) then follows from
\begin{eqnarray} \label{eq19}
\frac{P(\rho)}{k_BT}=\int\limits_0^\rho d\rho'\,S^{-1}(q=0,\rho')\,.
\end{eqnarray}
Various attempts have been made to develop accurate theories for the equation of 
state of fluids consisting of nonspherical particles:

(a) Scaled particle theory, which is very successful for hard sphere fluids, has 
been extended to prolate and oblate ellipsoids of revolution \cite{cott:79}, however 
with moderate success when gauged against Monte Carlo simulations \cite{muld:85}. 
Recently, it has been shown \cite{over:05a} that the results of scaled particle 
theory \cite{savi:81,boub:75,over:05a} for platelike particles or the closely related 
model of hard cut spheres are in disagreement with computer simulation data 
\cite{fren:82,eppe:84,veer:92,bate:99,zhan:02a,zhan:02b,beek:04}.

(b) Onsager theory \cite{onsa:42,onsa:49}, based on the second virial coefficient 
alone, can be ''rescaled'' \cite{pars:79,lee:87,lee:89}. Although this semi-empirical 
procedure leads to reasonably good results for rodlike particles, it is much less 
satisfactory for platelike particles \cite{wens:04}.

(c) Many theoretical studies on hard sphere fluids and depletion agents use the 
so-called free volume theory \cite{lekk:92}, in which the free volume accessible 
to a single particle plays a major role. This free volume theory has 
been studied within a fundamental measure theory \cite{over:05a}. However, 
it has been demonstrated that the resulting third virial coefficients of the 
equation of state for both hard cylinders and hard cut spheres differ from 
computer simulation results (see tables I and II in Ref.~\cite{over:05a}). 
Theoretical approaches based on fundamental measure theory do not yield correct 
third virial coefficients and equation of states due to the occurrence 
of so-called lost cases, i.e., the fact that configurations of three particles 
with pairwise overlap but no triple overlap do not contribute to thermodynamic 
properties (see, e.g., Refs.~\cite{rose:88,rose:98,tara:97}).

These theoretical and computer simulation studies demonstrate that the understanding 
of thermodynamic properties of nonspherical particles needs to be improved. Recently,
the measured inverse structure factor $S^{-1}(q=0,\rho)$ extrapolated to vanishing 
scattering vectors of stilbenoid dendrimers of the third generation has been modelled
in terms of the so-called $y3$ expansion \cite{barb:79,barb:80}
\begin{eqnarray}
\lefteqn{S^{-1}(q=0,\rho)=} \nonumber
\\&&\!\!\!\!\!\!\frac{1+2(B_2V_p-2)\phi+(3B_3V_p^2-8B_2V_p+6)\phi^2}{(1-\phi)^4} \label{eq20}
\\&\approx&1+2B_2\rho+3B_3\rho^2+O(\rho^3)\,.\label{eq21}
\end{eqnarray}
The $y3$ theory reproduces the exact second and third virial coefficients, 
$B_2$ and $B_3$, respectively. However, its practical applicability is limited 
due to the difficult numerical evaluation of the third virial coefficient
$B_3$ in the case of nonspherical particles \cite{harn:02b}, while the second 
virial coefficient $B_2$ for an isotropic hard convex body fluid is known exactly 
(see Ref.~\cite{harn:06} and references therein):
\begin{eqnarray} \label{eq22}
B_2=V_p+A_p\tilde{R_p}\,.
\end{eqnarray}
Here $A_p$ and ${\tilde R_p}=(1/4\pi)\int dA_p\,H_p$ are the surface area and the 
mean radius, respectively, where the local mean curvature is denoted as $H_p$. For a 
circular platelet of radius $R$ and thickness $L$ one has $V_p=\pi R^2L$, 
$A_p=2\pi R(R+L)$, and ${\tilde R}_p=\pi R/4+L/4$. For platelike stilbenoid dendrimers 
of the third generation the second virial coefficient $B_2/V_p=5.54$ as calculated from 
Eq.~(\ref{eq22}) with \mbox{$R=2.4$ nm} and \mbox{$L=1.8$ nm} agrees with the 
experimentally determined second virial coefficient \cite{harn:07}.

\begin{figure}[t!]
\begin{center}
\includegraphics[width=8cm, clip=]{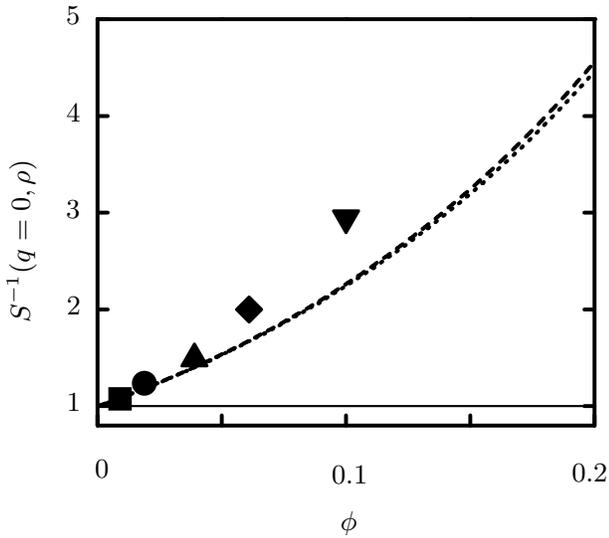}
\caption{Inverse structure factor $S^{-1}(q=0,\rho)$ extrapolated to vanishing 
scattering vectors of stilbenoid dendrimers of the third generation 
(with the same symbol code as in Fig.~\ref{fig5}). The dashed line follows from 
the scaled particle theory according to Eqs.~(\ref{eq20}), (\ref{eq22}), and (\ref{eq23}) 
while the dotted line represents the results of the fundamental measure theory as 
obtained from Eqs.~(\ref{eq20}), (\ref{eq22}), and (\ref{eq24}) \cite{harn:07}.}
\label{fig9}
\end{center} 
\end{figure}

In the framework of both scaled particle theory and Rosenfeld's fundamental 
measure theory $S^{-1}(q=0,\rho)$ is also given by Eqs.~(\ref{eq20}) and (\ref{eq21}) 
but the third virial coefficient reads
\begin{eqnarray} \label{eq23}
B_3^{(SPT)}=V_p^2+2{\tilde R}_pA_pV_p+\frac{1}{3}{\tilde R}_p^2A_p^2
\end{eqnarray}
within scaled particle theory and
\begin{eqnarray} \label{eq24}
B_3^{(FMT)}=V_p^2+2{\tilde R}_pA_pV_p+\frac{1}{12\pi}A_p^3
\end{eqnarray}
within fundamental measure theory. In Fig.~\ref{fig9} the experimentally determined 
inverse structure factor \mbox{$S^{-1}(q=0,\rho)$} extrapolated to vanishing scattering 
vectors of stilbenoid dendrimers of the third generation is compared with the results 
of scaled particle theory according to Eqs.~(\ref{eq20}), (\ref{eq22}), and (\ref{eq23}) 
and fundamental measure theory according to Eqs.~(\ref{eq20}), (\ref{eq22}), and (\ref{eq24}). 
With increasing volume fraction the theoretical results of both scaled particle theory 
(dashed line) and fundamental measure theory (dotted line) deviate from the experimental 
data (symbols). These deviations are mainly due to the fact that the predicted third 
virial coefficients $B_3^{(SPT)}$ and $B_3^{(FMT)}$ are too small. However, the PRISM
theory discussed above leads to a agreement with the experimental data as is apparent 
from Fig.~\ref{fig5} \cite{harn:07}. 

%
%
%
%

This comparison of experimentally determined $S^{-1}(q=0,\rho)$ of platelike particles 
with the predictions of the well-known scaled particle and fundamental measure theory 
has confirmed earlier caveats concerning the applicability of these theories to freely 
rotating nonspherical particles \cite{over:05a}. For example, 
$B_3^{(FMT)}=2 \pi^2 R^6/3 = 6.580 R^6$ as calculated from Eq.~(\ref{eq24}) for thin 
circular platelets in the limit $L \to 0$ underestimates the third virial coefficient 
$B_3 = 10.829 R^6$ according to computer simulation studies \cite{eppe:84,bate:99}.

\section{Orientational order in spatially homogeneous bulk phases} 
In this section we will outline some points on the basic description of orientational order 
in platelet dispersions and we will record details about the phase behavior of spatially
homogeneous bulk fluids. 

\subsection{Density functional theory}

\begin{figure}[t!]
\begin{center}
\includegraphics[width=4cm, clip=]{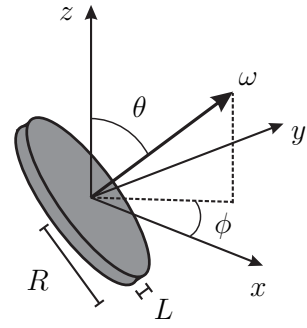}
\caption{Illustrations of the polar angle $\theta$ and the azimuthal angle $\phi$ of 
the normal \mbox{\boldmath$\omega$} of a platelet of radius $R$ and thickness $L$.
The center of mass of the particle is located at ${\bf r}=(0,0,0)$.} 
\label{fig10}
\end{center}
\end{figure}

The systems under investigation are dispersions involving $N$ species of colloidal 
particles. The number density of the centers of mass of platelets of species $i$ at 
a  point ${\bf r}$ with an orientation $\mbox{\boldmath$\omega$}=(\theta,\phi)$ of the 
normal of  the platelets (see Fig.~\ref{fig10}) is denoted by 
$\rho_i({\bf r},\mbox{\boldmath$\omega$})$ while for 
spherical structureless particles the center of mass number density is independent of
$\mbox{\boldmath$\omega$}$. The equilibrium density profiles of the fluid under the 
influence of external potentials $U_i({\bf r},\mbox{\boldmath$\omega$})$ minimize the 
grand potential functional \cite{evan:92,wu:06,harn:06}
\begin{eqnarray} \label{eq26} 
\Omega[\{\rho_i\}]&=&F_{id}[\{\rho_i\}]+F_{ex}[\{\rho_i\}]\nonumber
\\&+&\!\!\sum\limits_{i=1}^N \int d^3 r\,d\mbox{\boldmath$\omega$}\,
\rho_i({\bf r},\mbox{\boldmath$\omega$})
\left(U_i({\bf r},\mbox{\boldmath$\omega$})-\mu_i\right).
\end{eqnarray} 
Here $\mu_i$ is the chemical potential of particles of species $i$ and
$F_{id}[\{\rho_i\}]+F_{ex}[\{\rho_i\}]$ is the intrinsic Helmholtz free energy of the 
fluid (i.e., the Helmholtz free energy in the absence of the external potentials).
$F_{id}[\{\rho_i\}]$ is the intrinsic Helmholtz free energy of an ideal gas of 
noninteracting particles, 
\begin{eqnarray} \label{eq27} 
F_{id}[\{\rho_i\}]&=&k_BT\sum\limits_{i=1}^N 
\int d^3 r\,d\mbox{\boldmath$\omega$}\,\rho_i({\bf r},\mbox{\boldmath$\omega$})\nonumber
\\&&\times 
\left(\ln[4\pi\Lambda_i^3\rho_i({\bf r},\mbox{\boldmath$\omega$})]-1\right)\,,
\end{eqnarray}
where $\Lambda_i$ is the corresponding thermal de Broglie wavelength.
The excess (over the ideal gas) free energy functional
$F_{ex}[\{\rho_i\}]$ is in general a very complicated, highly non-trivial object,
because it is a characteristic property of an interacting many-body system. The 
functional $F_{ex}[\{\rho_i\}]$ is dealt with in various ways, which specify the 
explicit forms of the theory.

%
%
%
%

By means of a diagrammatic expansion it can be shown that the exact excess free 
energy functional in the low density limit reduces to a virial approximation, which 
is the starting point of the following considerations. Within a third-order virial 
approximation the excess free energy functional is given by
\begin{eqnarray} \label{eq28} 
\lefteqn{F_{ex}[\{\rho_i\}]=}\nonumber 
\\&&-\frac{k_BT}{2}\sum\limits_{i,j=1}^N 
\int d^3 r_1\,d^3 r_2\,d\mbox{\boldmath$\omega$}_1\,d\mbox{\boldmath$\omega$}_2\, 
\rho_i({\bf r}_1,\mbox{\boldmath$\omega$}_1)\nonumber
\\&&\times f_{ij}({\bf r}_1,\mbox{\boldmath$\omega$}_1;{\bf r}_2,\mbox{\boldmath$\omega$}_2) 
\rho_j({\bf r}_2,\mbox{\boldmath$\omega$}_2)\nonumber 
\\&&\times (1+\frac{1}{3}\sum\limits_{k=1}^N\int  d^3 r_3\,d\mbox{\boldmath$\omega$}_3\, 
f_{jk}({\bf r}_2,\mbox{\boldmath$\omega$}_2;{\bf r}_3,\mbox{\boldmath$\omega$}_2) \nonumber
\\&& \times \rho_k({\bf r}_3,\mbox{\boldmath$\omega$}_3) 
f_{ki}({\bf r}_3,\mbox{\boldmath$\omega$}_3;{\bf r}_1,\mbox{\boldmath$\omega$}_1))\,, 
\end{eqnarray} 
where $f_{ij}({\bf r}_1,\mbox{\boldmath$\omega$}_1;{\bf r}_2,\mbox{\boldmath$\omega$}_2)$ 
is the Mayer function of 
the pair interaction potential between two particles of species $i$ and $j$.
For sterically stabilized colloidal particles the most important feature of their
pair interaction is the strong mutual repulsion whenever the distance between 
the centers of mass of the particles is smaller than that at which their
surfaces touch each other. For nonspherical particles this shape-dependent 
repulsion is a function of the mutual orientation of the particles. In view of 
the steepness of the repulsive potential between two particles, the latter may 
be regarded  as hard particles, such that the pair interaction potential is 
infinite if the particle volumes overlap and is zero otherwise. By matching the
indices of refraction of the colloidal particles and the solvent it is  possible 
to effectively switch off the omnipresent dispersion forces and to create colloidal 
suspensions in which the actual effective interaction  between the colloidal 
particles very closely resembles this hard core potential. In this way theoretical 
models of hard body fluids have their  actual experimental counterparts with which 
their properties can be compared  quantitatively. For such hard body fluids the 
corresponding Mayer function equals $-1$ if the hard particles overlap and is 
zero otherwise \cite{hans:86}. Two thin platelets ($R\gg L$, see Fig.~\ref{fig10}), 
separated by a distance ${\bf r}_{12}$, intersect if the inequality  
\begin{eqnarray} \label{eq29} 
\left|{\bf r}_{12} \cdot 
(\mbox{\boldmath$\omega$}_1 \times \mbox{\boldmath$\omega$}_2)\right| 
&<&\sqrt{R^2\sin^2\!\gamma_{12}- 
({\bf r}_{12} \cdot \mbox{\boldmath$\omega$}_1)^2} \nonumber
\\&+&\sqrt{R^2\sin^2\!\gamma_{12}- 
({\bf r}_{12} \cdot \mbox{\boldmath$\omega$}_2)^2} 
\end{eqnarray}  
with $\mbox{\boldmath$\omega$}_j=(\sin\theta_j\cos\phi_j,\sin\theta_j 
\sin\phi_j,\cos\theta_j)\,,\; j=1,2$, and ${\bf r}_{12}={\bf r}_1-{\bf r}_2$ 
is fulfilled \cite{eppe:84}. $\gamma_{12}$ is the angle between the  
normals $\mbox{\boldmath$\omega$}_1$ and $\mbox{\boldmath$\omega$}_2$ of the  
two platelets. For thin platelets the intersection volume is like a line 
segment for $\mbox{\boldmath$\omega$}_1 \neq \mbox{\boldmath$\omega$}_2$.
Configurations with exactly equal orientations $\mbox{\boldmath$\omega$}_1 =
\mbox{\boldmath$\omega$}_2$ are not considered for freely rotating thin platelets 
in three dimensions because they have a vanishing statistical weight.
Minimization of $\Omega$ with respect to $\rho_i({\bf r},\mbox{\boldmath$\omega$})$ 
leads to integral equations for the equilibrium density profiles:
\begin{eqnarray} \label{eq30} 
\rho_i({\bf r},\mbox{\boldmath$\omega$})&=&
\frac{1}{4\pi \Lambda_i^3}\exp\left(-\frac{1}{k_BT}
\frac{\partial F[\{\rho_i\}]}{\partial \rho_i({\bf r},\mbox{\boldmath$\omega$})}\right)\,.
\end{eqnarray} 
These equations can be solved numerically for given chemical potentials $\mu_i$ 
and given external potentials $U_i({\bf r},\mbox{\boldmath$\omega$})$ using a Picard 
scheme with retardation, i.e., the equations 
$\rho_i({\bf r},\mbox{\boldmath$\omega$})=J[\{\rho_i\}]$ are solved according to  
$\rho_i^{(l)}({\bf r},\mbox{\boldmath$\omega$})=mix\, J[\{\rho_i^{(l-1)}\}]+ 
(1-mix)\rho_i^{(l-1)}({\bf r},\mbox{\boldmath$\omega$})$ with a mixing parameter 
$0< mix < 1$. Starting from an initial guess for 
$\rho_i^{(0)}({\bf r},\mbox{\boldmath$\omega$})$, the solutions of 
the equations are obtained by an iterative approach $l=0, 1, 2, ...\,$. The mixing  
parameter is adjusted empirically to ensure convergence. In practice, first the Mayer  
functions are calculated and are stored for all required values of  
$({\bf r}_1,\mbox{\boldmath$\omega$}_1;{\bf r}_2,\mbox{\boldmath$\omega$}_2)$. 
Thereafter the integral equations [Eq.~(\ref{eq30})] are solved.

For a homogeneous and isotropic bulk fluid with $U_i({\bf r},\mbox{\boldmath$\omega$})=0$ in a 
macroscopic volume $V$ the grand potential functional [Eq.~(\ref{eq26})] reduces to
\begin{eqnarray} \label{eq31} 
\frac{\Omega_b}{Vk_BT}&=&\sum\limits_{i=1}^N\rho_i 
\left(\ln(\Lambda_i^3\rho_i)-1-\mu_i(k_BT)^{-1}\right)\nonumber
\\&+&\sum\limits_{i,j=1}^N\rho_i\rho_j\left(B^{(ij)}_2 
+\frac{1}{2}\sum\limits_{k=1}^NB^{(ijk)}_3\rho_k\right)\,, 
\end{eqnarray} 
where $\rho_i=V^{-1}\int d^3 r\,d\mbox{\boldmath$\omega$}\,\rho_i({\bf r},\omega)$ are 
the total particle number densities. $B^{(ij)}_2$ is the second virial coefficient between 
particles of species $i$ and $j$: 
\begin{equation} \label{eq32} 
B^{(ij)}_2=-\frac{1}{2(4\pi)^2}\int d^3 r_1\,d\mbox{\boldmath$\omega$}_1
\,d\mbox{\boldmath$\omega$}_2\,
f_{ij}({\bf r}_1,\mbox{\boldmath$\omega$}_1;0,\mbox{\boldmath$\omega$}_2)\,.
\end{equation} 
In accordance with the standard definition \cite{hans:86}, the third virial coefficient
involving three particles of species $i$, $j$, $k$ is given by  
\begin{eqnarray} \label{eq33} 
B^{(ijk)}_3&=&-\frac{1}{3(4\pi)^3}\int d^3 r_1\,d^3 r_2\, 
d\mbox{\boldmath$\omega$}_1\,d\mbox{\boldmath$\omega$}_2\,d\mbox{\boldmath$\omega$}_3\nonumber 
\\&& \times f_{ij}({\bf r}_1,\mbox{\boldmath$\omega$}_1;0,\mbox{\boldmath$\omega$}_2)
f_{jk}({\bf r}_1-{\bf r}_2,\mbox{\boldmath$\omega$}_2;0,\mbox{\boldmath$\omega$}_3)\nonumber
\\&& \times f_{ki}({\bf r}_2,\mbox{\boldmath$\omega$}_3;0,\mbox{\boldmath$\omega$}_1)\,. 
\end{eqnarray} 
The equation of state derived from the grand potential [Eq.~(\ref{eq31})] takes the 
following form: 
\begin{eqnarray} 
\frac{P}{k_BT}&=&-\frac{1}{k_BT}
\left(\frac{\partial \Omega_b}{\partial V}\right)_{T,\{ \mu_i \}}
\\&=&\left(\sum\limits_{i=1}^N\rho_i 
+\sum\limits_{i,j=1}^NB_{ij}^{(2)}\rho_i\rho_j 
+\sum\limits_{i,j,k=1}^NB_{ijk}^{(3)}\rho_i\rho_j\rho_k\right)\,.\nonumber
\\&&\label{eq34} 
\end{eqnarray} 
The same equation without the second and the third term on the right-hand side holds  
for the ideal gas limit, i.e., noninteracting particles.

\subsection{Fundamental measure theory for thin hard platelets in three dimensions}
\begin{figure}[t!]
\begin{center}
\includegraphics[width=8cm, clip=]{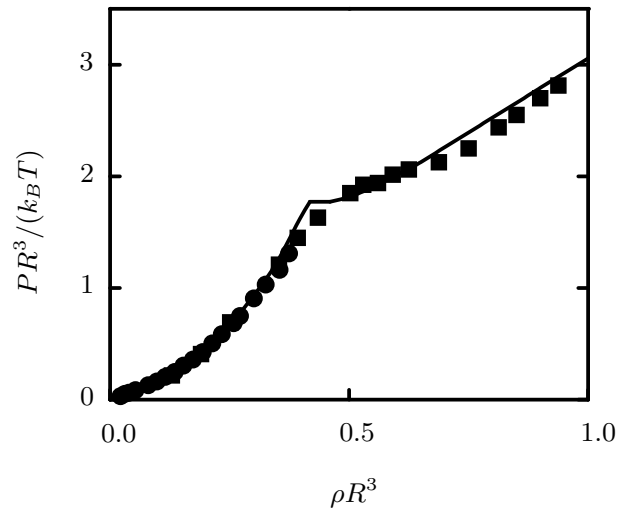}
\caption{Equation of state of thin platelets ($R\gg L$, see Fig.~\ref{fig10}) as obtained 
from fundamental measure theory according to Eqs.~(\ref{eq35}) - (\ref{eq44}) (solid line 
\cite{eszt:06}) and from  computer simulations (circles \protect\cite{eppe:84} and squares 
\protect\cite{dijk:97}). The horizontal line is the tie-line illustrating 
isotropic-nematic two-phase coexistence according to the fundamental 
measure theory.}
\label{fig11}
\end{center} 
\end{figure}

Within a geometry-based density functional theory the excess free energy 
functional for monodisperse thin platelets in three dimensions is obtained by integrating 
over an excess free energy density, 
\begin{equation} \label{eq35} 
F_{ex}[\rho]=\frac{k_BT}{16\pi^2}\int d^3 r\,
d \mbox{\boldmath$\omega$}_1\,d \mbox{\boldmath$\omega$}_2\,\Phi(\{n_\nu\})\,,
\end{equation} 
where the spatial and angular arguments of the weighted densities $n_\nu$ are 
suppressed in the notation. The free energy density $\Phi$ is given by 
\cite{eszt:06}
\begin{eqnarray} \label{eq36} 
\Phi(\{n_\nu\})&=&n_1({\bf r},\mbox{\boldmath$\omega$}_1)
n_2({\bf r},\mbox{\boldmath$\omega$}_1)\nonumber
\\&+&\frac{1}{24 \pi}n_2({\bf r},\mbox{\boldmath$\omega$}_1)
n_3({\bf r},\mbox{\boldmath$\omega$}_1;\mbox{\boldmath$\omega$}_2)
n_2({\bf r},\mbox{\boldmath$\omega$}_2). \label{eq37}
\end{eqnarray} 
The weighted densities are related to the number density of the center 
of mass of the platelets $\rho({\bf r},\mbox{\boldmath$\omega$})$ according to
\begin{eqnarray} 
n_1({\bf r},\mbox{\boldmath$\omega$}_1)\!\!\!&=&
\!\!\!\!\int\frac{d \mbox{\boldmath$\omega$}'}{4\pi}
w_1({\bf r},\mbox{\boldmath$\omega$}';\mbox{\boldmath$\omega$}_1)
\star\rho({\bf r},\mbox{\boldmath$\omega$}')\,,  \label{eq38}
\\n_2({\bf r},\mbox{\boldmath$\omega$}_1)\!\!\!&=&
w_2({\bf r},\mbox{\boldmath$\omega$}_1)
\star\rho({\bf r},\mbox{\boldmath$\omega$}_1)\,,    \label{eq39}
\\n_3({\bf r},\mbox{\boldmath$\omega$}_1;\mbox{\boldmath$\omega$}_2)\!\!\!&=&
\!\!\!\!\int\frac{d \mbox{\boldmath$\omega$}'}{4\pi}
w_3({\bf r},\mbox{\boldmath$\omega$}';\mbox{\boldmath$\omega$}_1;\mbox{\boldmath$\omega$}_2)
\star\rho({\bf r},\mbox{\boldmath$\omega$}'),   \label{eq40}
\end{eqnarray} 
where the asterisk $\star$ denotes the spatial convolution and the weight functions 
are given by
\begin{eqnarray} 
w_0({\bf r},\mbox{\boldmath$\omega$}_1)&=&\!\!\!
\frac{1}{8}\delta(R-|{\bf r}|)\delta({\bf r}\cdot\mbox{\boldmath$\omega$}_1)\,,  \label{eq41}
\\w_2({\bf r},\mbox{\boldmath$\omega$}_1)&=&\!\!\!
2\Theta(R-|{\bf r}|)\delta({\bf r}\cdot\mbox{\boldmath$\omega$}_1)\,,            \label{eq42}
\\w_1({\bf r},\mbox{\boldmath$\omega$}_1;\mbox{\boldmath$\omega$}_2)&=&\!\!\!
\frac{2}{R}|\mbox{\boldmath$\omega$}_1\cdot
(\mbox{\boldmath$\omega$}_2  \times {\bf r})|
w_0({\bf r},\mbox{\boldmath$\omega$}_1)\,,                                        \label{eq43}
\\w_3({\bf r},\mbox{\boldmath$\omega$}_1;\mbox{\boldmath$\omega$}_2;\mbox{\boldmath$\omega$}_3)
&=&\!\!\!
\frac{8}{\pi}|\mbox{\boldmath$\omega$}_1\cdot
(\mbox{\boldmath$\omega$}_2 \times \mbox{\boldmath$\omega$}_3)|
w_2({\bf r},\mbox{\boldmath$\omega$}_1).                                     \label{eq44}
\end{eqnarray} 
Here $\Theta(x)$ is the Heaviside step function and $\delta(x)$ is the Dirac 
distribution. The Mayer function of the pair interaction potential between two hard 
platelets [Eq.~(\ref{eq29})] is recovered via the convolution
\begin{eqnarray} \label{eq45}
f({\bf r},\mbox{\boldmath$\omega$}_1;0,\mbox{\boldmath$\omega$}_2)=
-2w_2({\bf r},\mbox{\boldmath$\omega$}_1)\star
w_1({\bf r},\mbox{\boldmath$\omega$}_1;\mbox{\boldmath$\omega$}_2)\,.
\end{eqnarray}
However, the fundamental measure theory does not yield the correct third virial coefficient 
due to the above mentioned occurrence of lost cases \cite{eszt:06}. The fundamental measure 
approximation for the third virial coefficient is nonvanishing only for three platelets at 
position vectors ${\bf r}_1, {\bf r}_2, {\bf r}_3$ with a common triple 
intersection. The lost cases are compensated by overcounting the cases with triple intersections, 
such that the resulting equation of state is in reasonable agreement with the exact results 
for the isotropic bulk fluid. In the case of thin platelets this has been achieved by using 
the prefactor $1/(24 \pi)$ in the second term in Eq.~(\ref{eq36}) which leads to the 
third virial coefficient $B_3^{(FMT)}=2\pi^2 R^6/3$. The equation of state 
as obtained from this fundamental measure theory is indeed in reasonable agreement with computer 
simulation data as is apparent from Fig.~\ref{fig11}. 

The density functional theory allows one to consider the possibility of an isotropic 
to nematic phase transition in a monodisperse fluid of platelets as a function of particle 
number density. The broken symmetry in the spatially homogeneous nematic phase may be 
characterized by the nematic order parameter 
\begin{eqnarray} \label{eq46}
S=\frac{1}{8 \pi\rho V}\int d^3 r\,d \mbox{\boldmath$\omega$}
\rho({\bf r},\mbox{\boldmath$\omega$})(3\cos^2\theta -1)\,
\end{eqnarray}
which varies between $0$ in the isotropic phase and $1$ when all platelets are 
oriented along one axis. For the bulk isotropic (I) to nematic (N) phase transition the 
following values of the particle number density and the nematic order parameter 
at two-phase coexistence have been obtained from fundamental measure theory:
$\rho_I R^3=0.42$, $\rho_N R^3=0.46$, $S_I=0.05$, and $S_N=0.49$ \cite{eszt:06}. These values 
are similar to the ones obtained from computer simulations while a second virial approximation 
based on Eq.~(\ref{eq28}) without the last term overestimates both the density 
jump and the nematic order parameter at coexistence and yields 
$\rho_I R^3=0.67$, $\rho_N R^3=0.85$, and $S_N=0.79$. Even for thin platelets the chance 
that one platelet has contact with two or more platelets simultaneously is significant, 
and thus for this fluid the third virial coefficient is non-negligible. For comparison
we note that the contribution of the third virial coefficient to the equation of state 
of a fluid consisting of thin rods is vanishing small, reflecting the small probability 
that three rods will simultaneously intersect \cite{onsa:49}.
%
%
%
%

Density functional theory has been applied to fluids consisting of thin hard platelets
in contact with a hard substrate \cite{harn:02a}. Platelets lying very close to the 
substrate must adopt nearly a fully parallel alignment due to interactions with the substrate.
The probability of finding platelets touching the substrate is increased compared with the 
bulk due to effective entropic interactions. Upon increasing the bulk platelet density 
the substrate is completely wetted by a nematic film \cite{reic:07a}. For the fluid
confined by two parallel hard substrates, a first-order capillary nematization transition 
for large slit widths has been determined \cite{reic:07}. 

The enrichment and ordering of clay platelets near surfaces is important 
for the oil and gas production \cite{mait:00}. The first stage of recovering oil and gas 
from an underground reservoir is the drilling of wells into the hydrocarbon-bearing rock. 
This is performed by rotary drilling using, for example, water-based drilling fluids that
contain smectic clay platelets such as bentonite or montmorillonite. Among the many 
functions that drilling fluids must possess to be successful for a drilling operation are: 
carrying the drilling cuttings and transporting them back to the surface; suspending the 
drilled cuttings when circulation is stopped; cooling and cleaning the bit; reducing the 
friction between the drilling string and the sides of the hole; preventing the inflow of 
fluids from the permeable rocks that are penetrated; maintaining the stability of the 
wellbore; and forming a thin, low permeability filter-cake which seals the pores and other 
openings in formations penetrated by the bit. Scanning electron microscope images of 
mixed-metal hydroxide - bentonite filter-cake formed on the surface of a sandstone 
rock show in detail how the mixed-metal hydroxide and bentonite platelets form bridges 
across the entrance of rock pores of size \mbox{30 $\mu$m} and larger, although the radius 
of an individual platelet is less than \mbox{0.5 $\mu$m} \cite{mait:00}. Effective entropic 
interactions together with van der Waals interactions at submicron distances play an 
important role in understanding the underlying mechanisms of orientational ordering 
near surfaces \cite{perd:05}. Density functional theory provides the basic description of 
such inhomogeneous colloidal platelet fluids \cite{harn:06}.

\subsection{Strictly two-dimensional hard platelets}
\begin{figure}[t!]
\begin{center}
\includegraphics[width=8cm, clip=]{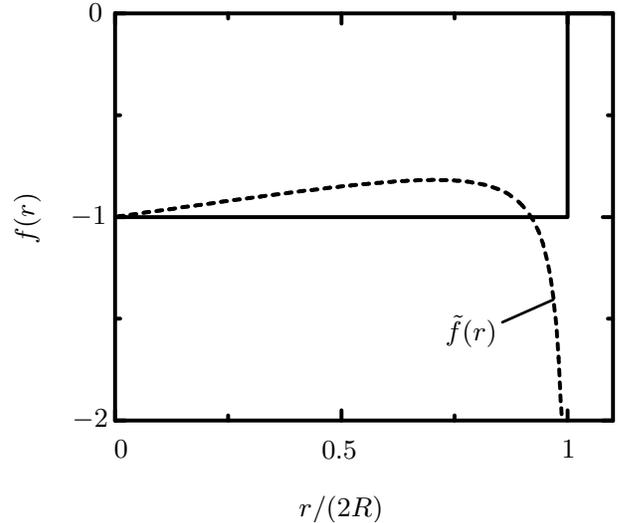}
\caption{The exact Mayer function $f(r)$ for platelets of radius $R$ in two dimensions 
(solid line) and the approximate representation ${\tilde f}(r)$ according to the 
fundamental measure theory (dashed line) as obtained from Eq.~(\ref{eq52}), respectively.
The distance between the centers of the platelets is denoted by $r$. Note that 
${\tilde f}(r)$ diverges upon approaching $r=2R$.}
\label{fig12}
\end{center} 
\end{figure}

The density functional theory presented in the previous subsection is based 
on the weight functions given by Eqs.~(\ref{eq41}) - (\ref{eq44}) and the deconvolution 
of the Mayer function [Eq.~(\ref{eq45})]. Although the particle shape is equivalent,
overlapping pair configurations are very different for freely rotating platelets 
in three dimension as compared to the strictly two-dimensional case. While for thin 
platelets in three dimensions the intersection volume is like a line segment 
($\mbox{\boldmath$\omega$}_1 \neq \mbox{\boldmath$\omega$}_2$), the intersection 
volume is an area for platelets in two dimensions ($\mbox{\boldmath$\omega$}_1 = \mbox{\boldmath$\omega$}_2$). Rosenfeld's weight functions for two-dimensional hard 
platelets of radius $R$ are \cite{rose:94,rose:95,rose:96,tara:97}
\begin{eqnarray}  \label{eq47}
w_2({\bf r})&=&\Theta(R-|{\bf r}|)\,,
\\w_1({\bf r})&=&\delta(R-|{\bf r}|)\,,  \label{eq48}
\\{\bf w}_1({\bf r})&=&w_2({\bf r})\frac{{\bf r}}{|{\bf r}|}\,,  \label{eq49}
\\w_0({\bf r})&=&w_1({\bf r})\frac{1}{2\pi R}\,. \label{eq50}
\end{eqnarray}
The exact Mayer function is approximated by 
\begin{eqnarray}  \label{eq51}
{\tilde f}({\bf r})&=&2 w_0({\bf r}) \star w_2({\bf r})\nonumber
\\&&+\frac{1}{2\pi}\left(w_1({\bf r}) \star w_1({\bf r})-
{\bf w}_1({\bf r}) \star {\bf w}_1({\bf r}) \right)\,,
\end{eqnarray}
where ${\bf r}$ is the center to center vector and the asterisk $\star$ denotes 
the two-dimensional convolution. The evaluation of the integrals in Eq.~(\ref{eq51}) 
yields \cite{eszt:06}
\begin{eqnarray}  \label{eq52}
\lefteqn{{\tilde f}(r)=} \nonumber
\\&&\!\!\!\Theta(2R-r)\left(\frac{2}{\pi}\arccos\left(\frac{r}{2R}\right)+
\frac{r}{\pi \sqrt{4R^2-r^3}}\right).
\end{eqnarray}
Figure \ref{fig12} illustrates that $f({\bf r})$ and 
${\tilde f}({\bf r})$ differ qualitatively. In particular, ${\tilde f}({\bf r})$
exhibits a divergence in the case that the two platelets touch each other, i.e., 
for $|{\bf r}|=2 R$. However the deviations from the exact value (solid line in 
Fig.~\ref{fig12}) balance such that the second virial coefficient is exact, i.e., 
$\int d^2r\, f({\bf r})= \int d^2r\, {\tilde f}({\bf r})$.
Higher order virial coefficients have been calculated recently using molecular
dynamics simulations \cite{kola:06}. Moreover, the fundamental measure theory for 
two-dimensional platelets has been considered as a useful tool to study inhomogeneous 
systems such as laser induced freezing and melting of confined colloidal particles 
\cite{rasm:02} despite the deficiencies of the theory in reproducing the Mayer 
function.
%
%
%
%

\subsection{Colloidal mixtures of hard spheres and platelets}
Depletion interactions between large colloidal particles induced by smaller particles, 
which can be either solvent particles or a collodial component in its own right, are of
significant research interest because of the importance of these effective interactions
in various colloidal processes. For example, flocculation of colloids can be driven by 
the addition of non-adsorbing polymers via the depletion mechanism. On the other hand, 
stabilization of colloidal suspensions can be achieved by the addition of adsorbing 
polymers. Whereas experimental and theoretical studies have focused on binary 
hard-sphere fluids as well as on colloidal mixtures of hard spheres and hard rods 
or polymers, less attention has been paid to hard platelets acting as depletants
\cite{piec:00,over:03,harn:04a,over:04a,over:04b,over:05,harn:05a,over:05a}.

Density functional theory for thin hard platelets has been used to investigate 
the depletion potential between two hard spheres due to the presence of thin platelets 
\cite{harn:04a}. The calculated depletion potential between two spheres exhibits an 
attractive primary minimum at contact when the spheres touch each other. This minimum 
deepens upon increasing the platelet density and a small repulsive barrier at larger 
sphere separations develops. However, the depletion barrier is typically less than the 
thermal energy $k_BT$, and therefore unlikely to significantly alter the kinetics 
of aggregation of the hard spheres at platelet densities smaller than one half the 
density of the isotropic phase at bulk isotropic-nematic two-phase coexistence. Nonetheless, 
with increasing platelet density the integrated strength of the effective interaction 
between the spheres becomes significantly weaker and thus reduces the thermodynamic 
onset of flocculation \cite{harn:04a}. 

\begin{figure}[t!]
\begin{center}
\includegraphics[width=8cm, clip=]{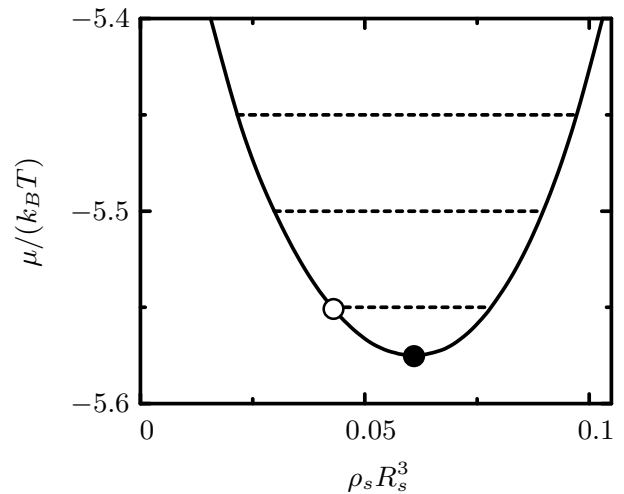}
\caption{Bulk and surface phase diagrams of binary mixtures of spheres of radius $R_s$ 
and thin platelets of radius $R=R_s/2$ as a function of the chemical potential of the 
platelets $\mu$ and the number density of the spheres $\rho_s$ \cite{harn:05a}.
The straight dashed lines are tie-lines illustrating liquid-liquid phase coexistence. 
The solid and open circles denote the bulk critical point and the wetting transition 
point, respectively. Between the wetting transition point and the critical point the 
sphere-rich liquid phase completely wets the interface between a hard substrate and 
the sphere-poor liquid phase.}
\label{fig13}
\end{center} 
\end{figure}

A fundamental measure-based density functional theory has been used to study binary 
colloidal fluids consisting of hard spheres and thin platelets in their bulk and 
near a planar hard substrate. Figure \ref{fig13} displays the calculated phase diagram
for a binary mixture of spheres and thin platelets for a size ratio $R_S/R=2$, where 
$R_s$ is the radius of the spheres \cite{harn:05a}. The tie-lines are horizontal 
because the coexisting phases are characterized by the equality of the chemical 
potential of the platelets $\mu$ in the coexisting phases. The binodal for coexisting 
states is shown, for which a sphere-rich and a platelet-poor liquid phase coexists 
with a sphere-poor and a platelet-rich liquid phase. The coexistence region is bounded
by a lower critical point below which only a single stable phase is found.
Upon decreasing $\mu$ the adsorption behavior of the mixture near a planar substrate 
changes qualitatively so that two cases have to be distinguished. For 
$\mu/(k_BT)>-5.551$ the sphere-rich phase does not wet the wall at two-phase coexistence.
The layer thickness of the sphere-rich phase forming close to the wall increases 
continuously upon approaching the bulk phase boundary, but remains finite at coexistence.
For $\mu/(k_BT)<-5.551$, however, the wall is wetted completely. The transition to 
complete wetting is first-order because the excess adsorptions of both the spheres 
and platelets along the bulk coexistence curve jump to macroscopic values upon 
crossing the wetting transition point.
%
%
%
%

Binary mixtures of silica spheres of radius \mbox{$R_s=350$ nm} and silica-coated 
gibbsite platelets of radius \mbox{$R=229$ nm} and  tickness \mbox{$L=56$ nm} have 
been investigated by means of confocal microscopy and TEM \cite{over:05}. The depletion-induced
aggregation of the spherical particles has been observed in a TEM micrograph. In contrast 
to sphere-polymer and sphere-rod mixtures, the addition of platelets to a solution of 
spheres close to the fluid-solid binodal of the latter slowed down early crystalline 
ordering of the spheres, which led to more grain  boundaries and powder-like structures. 
It has been argued that in  the two-phase region the disordered sediment structures 
are kinetically arrested,  which  prevents further crystallization. Moreover, fluid-like 
microphases of  platelets were trapped in the sediment which is due to the simultaneous 
sedimentation  of platelets and spheres as well as due to depletion interactions.  

It has been shown that spheres and platelets can be separated by adding a third species 
to the suspension \cite{maso:02}. The aggregation of platelets (\mbox{$R=0.5$ $\mu$m} 
and $L=0.2$ $\mu$m) and spheres (\mbox{$R_s=0.5$ $\mu$m}) of 1-eicosene wax in a 
dilute aqueous suspension has been induced by nanometer-sized spherical micelles 
of sodium dodecyl sulfate. Below the critical micelle concentration neither the 
platelets nor the spheres aggregate. Upon increasing the micelle concentration the 
platelets aggregate into columns due to effective entropic attraction while the 
spheres remain unaggregated. The columnar aggregates of platelets resemble the 
stacked-coin structures of red blood cells, known as rouleaux, which form when 
polymers are added to blood. Analogous to fractioned crystallization, the 
aggregated platelets can be repeatedly separated from the unaggregated spheres by 
gravitational creaming, yielding a shape-specific colloidal purification method. For 
high micelle concentrations not only the platelets aggregate but also the spheres 
aggregate with other spheres. It has been argued that the observed shape-selective 
colloidal separation might be applicable to a broad class of mixtures of spherical 
and nonspherical colloids such as viruses, organelles, proteins, and cells \cite{maso:02}.

Very recently it has been investigated how the surface roughness of colloidal 
platelets can affect the depletion attraction between them \cite{zhao:07}.
The depletion attraction between polymeric pentagonal platelets 
(pseudo-radius \mbox{$R\approx 1.5$ $\mu$m} and thickness $L=1$ $\mu$m) can be suppressed 
when the nanoscale surface asperity heights on the face of the platelets become 
larger than the size of nanometer-sized spherical micelles of sodium dodecyl sulfate
acting as depletion agent. In the opposite limit, the attraction reappears and 
columnar stacks of platelets are formed. Exploiting this, the site-specific 
roughness on only one side of the platelets has been increased in order to form 
aligned platelet dimers. The two smoother surfaces of two Janus platelets 
aggregate face-to-face, exposing the rougher surfaces that cannot aggregate.
The general problem of how surface roughness affects the strength and range of the 
depletion attraction is interesting and needs to be elucidated in future.

\subsection{Density functional theory for the Zwanzig model}
\begin{figure}[t!]
\begin{center}
\includegraphics[width=5cm, clip=]{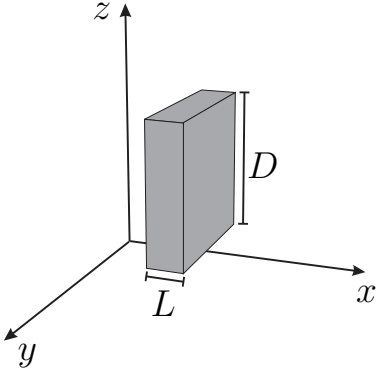}
\caption{Within the Zwanzig model, platelets are represented by square parallelepipeds 
of thickness $L$ and width $D$, which can take only three orientations, along the 
$x$, $y$ or $z$ directions. The number density of  the center of mass of the platelet 
in the figure is denoted by $\rho_x({\bf r})$.}
\label{fig14}
\end{center} 
\end{figure}

Taking into account the simultaneous presence of orientational ordering and
polydispersity in density functional calculations for freely rotating nonspherical 
particles is computationally demanding because the integrals to be evaluated in 
Eq.~(\ref{eq28}) are high-dimensional, already within a third-order virial approximation. 
In order to reduce this computational effort one may use the Zwanzig model for hard 
square parallelepipeds \cite{zwan:63}. Within the Zwanzig model particles of species 
$i=1, ..., N$ are represented by rectangular blocks of size $L_i\times D_i\times D_i$. 
The positions of the centers of mass vary continuously, while the orientations of the 
normal of each particle are restricted to discrete directions $\beta=x, y, z$ (see
Fig.~\ref{fig14}). Using the notation  
$\alpha_x^{(i)}(x,z)=\alpha_i(x,z,\theta=\pi/2,\phi=0)$,  
$\alpha_y^{(i)}(x,z)=\alpha_i(x,z,\theta=\pi/2,\phi=\pi/2)$, and 
$\alpha_z^{(i)}(x,z)=\alpha_i(x,z,\theta=0,\phi=0)$ for $\alpha=\rho, U$, 
the grand potential functional [Eq.~(\ref{eq26})] can be written as 
%
%
%
%
%
%
\begin{eqnarray}
\Omega[\{\rho^{(i)}_\beta({\bf r})\}]&=&F_{id}[\{\rho^{(i)}_\beta({\bf r})\}] 
+F_{ex}[\{\rho^{(i)}_\beta({\bf r})\}]\nonumber 
\\&+&\sum_{i=1}^N\sum_{\beta=x, y, z}\int d^3 r\, \rho^{(i)}_\beta({\bf r}) 
\left(U^{(i)}_\beta({\bf r})-\mu_i\right)\nonumber
\\ \label{eq53}
\end{eqnarray} 
with the ideal gas contribution [Eq.~(\ref{eq27})] 
\begin{eqnarray} \label{eq54}
\lefteqn{F_{id}[\{\rho^{(i)}_\beta({\bf r})\}]=}  \nonumber
\\&&k_BT\sum_{i=1}^N\sum_{\beta} 
\int d^3 r\,\rho^{(i)}_\beta({\bf r}) 
\left(\ln[\Lambda_i^3\rho^{(i)}_\beta({\bf r})]-1\right).
\end{eqnarray} 
The third-order virial approximation of the excess free energy functional
[Eq.~(\ref{eq28})] reads 
\begin{eqnarray} \label{eq55}
\lefteqn{F_{ex}[\{\rho^{(i)}_\beta({\bf r})\}]=}   \nonumber
\\&&-\frac{k_BT}{2}\!\!\sum_{i,j=1}^N 
\sum_{\beta_1,\beta_2}\!\int \!\! d^3 r_1 d^3 r_2
\rho^{(i)}_{\beta_1}({\bf r}_1) 
f^{(i j)}_{\beta_1 \beta_2}({\bf r}_1,{\bf r}_2) 
\rho^{(j)}_{\beta_2}({\bf r}_2)\nonumber 
\\&&\times (1+\frac{1}{3}\sum_{k=1}^2 
\sum_{\beta_3}\int d^3 r_3\, 
f^{(jk)}_{\beta_2\beta_3}({\bf r}_2,{\bf r}_3) \nonumber
\\&&\hspace*{3cm}\times \rho^{(k)}_{\beta_3}({\bf r}_3) 
f^{(ki)}_{\beta_3\beta_1}({\bf r}_3,{\bf r}_1))\,,
\end{eqnarray}
where $f^{(ij)}_{\beta_1\beta_2}({\bf r}_1,{\bf r}_2)$ is the Mayer function 
which for hard bodies as considered here equals $-1$ if two particles of species  
$i$ and $j$ with orientations $\beta_1$ and $\beta_2$  overlap and is zero otherwise.  
With the definition  
\begin{equation} \label{eq56} 
S_{\alpha\beta}^{(i)}=D_i+(L_i-D_i) \delta_{\alpha\beta}\,, 
\end{equation} 
which represents the spatial extent in direction $\alpha=x,y,z$ of a particle  
of species $i$ with orientation $\beta$ of the normal, the Mayer function 
can be written explicitly as  
\begin{eqnarray} \label{eq57}
\lefteqn{f^{(ij)}_{\beta_1\beta_2}({\bf r}_1,{\bf r}_2)=}   \nonumber
\\&&-\prod_{\alpha} 
\Theta\left(\frac{1}{2}\left(S_{\alpha\beta_1}^{(i)}+S_{\alpha\beta_2}^{(j)} 
\right)-|r_{\alpha 1}-r_{\alpha 2}|\right)\,, 
\end{eqnarray} 
where $r_{\alpha 1}$ is the projection of the position vector ${\bf r}_1$ in $\alpha$ 
direction. The density functional theory is completely specified by the excess free energy 
functional and the Mayer functions. Minimization of $\Omega$ with respect to 
$\rho_\beta^{(i)}({\bf r})$ leads to integral equations for the equilibrium density profiles.

Zwanzig's model may be considered as a coarse-grained version of the Onsager model, which 
allows for continuously varying orientations  \cite{onsa:42,onsa:49}. The model offers the
advantage that the determination of phase diagrams and density profiles  becomes numerically straightforward, allowing one  to study various aspects of platelet fluids in detail
and to scan wide parameter ranges. Moreover, Rosenfeld's very successful fundamental
measure theory for hard sphere systems has been extended to the Zwanzig model, which is an 
important alternative in view of the uncertain convergence behavior of the virial series of 
the equation of state of platelets of finite thickness \cite{harn:02c}. Within the 
fundamental measure theory, one postulates the following form of the excess free energy functional
\cite{cues:96,cues:97a,cues:97b}: 
\begin{equation} \label{eq58} 
F_{ex}[\{\rho^{(i)}_\beta({\bf r})\}]=k_BT \int d^3 r\,
\Phi(\{ \rho^{(i)}_\beta({\bf r}) \} )\,,
\end{equation} 
with the reduced excess free energy density
\begin{equation} \label{eq59} 
\Phi=-n_0\ln(1-n_3)+\frac{{\bf n}_1\cdot{\bf n}_2}{1-n_3}+ 
\frac{n_{2x}n_{2y}n_{2z}}{(1-n_3)^2}\,, 
\end{equation} 
and the weighted densities
\begin{equation} \label{eq60} 
n_l({\bf r})=\sum_{i=1}^N\sum_{\beta}\int d^3 r_1\, 
\omega_{l\,\beta}^{(i)}({\bf r}-{\bf r}_1)\rho^{(i)}_\beta({\bf r}_1) 
\end{equation} 
for $l\in\{0,1x, 1y, 1z, 2x, 2y, 2z, 3\}$. The weight functions
$\omega_{l\,\beta}^{(i)}({\bf r})$ are obtained by expressing the Fourier
transform of the Mayer function [Eq.~(\ref{eq57})] as a sum of products of single
particle functions. The explicit expressions of the weight functions read:
\begin{eqnarray} \label{eq61}
\omega_{0 \,\beta}^{(i)}({\bf r})&=&\frac{1}{8}\prod\limits_{\alpha=x,y,z}
\delta\left(S^{(i)}_{\alpha \beta}/2-|r_\beta|\right)\,,
\\\omega_{3\, \beta}^{(i)}({\bf r})&=&\prod\limits_{\alpha=x,y,z}
\Theta\left(S^{(i)}_{\alpha \beta}/2-|r_\beta|\right)\,,  \label{eq62}
\\\omega_{1 \alpha \,\beta}^{(i)}({\bf r})&=&
\frac{2\Theta\left(S^{(i)}_{\alpha \beta}/2-|r_\alpha|\right)}
{\delta\left(S^{(i)}_{\alpha \beta}/2-|r_\alpha|\right)}
\omega_{0\, \beta}^{(i)}({\bf r})\,,   \label{eq63}
\\\omega_{2 \alpha\,  \beta}^{(i)}({\bf r})&=&
\frac{\delta\left(S^{(i)}_{\alpha \beta}/2-|r_\alpha|\right)}
{2\Theta\left(S^{(i)}_{\alpha \beta}/2-|r_\alpha|\right)}
\omega_{3\, \beta}^{(i)}({\bf r})\,.  \label{eq64}
\end{eqnarray}
In the limit of monodisperse thin platelets ($L \ll D$, see Fig.~\ref{fig14}) the 
excess free energy functional [Eqs.~(\ref{eq58}) - (\ref{eq64})] reduces to a third-order 
virial approximation. It is worthwhile to emphasize that the fundamental measure theory 
for the Zwanzig model yields the exact second and third virial coefficients 
irrespective of the shape and size polydispersity of the particles. For monodisperse 
platelets the second virial coefficient is given by Eq.~(\ref{eq22}) with $V_p=LD^2$,
$A_p=2D(D+L)$, and ${\tilde R}_p=(L+2D)/3$, while the third virial coefficient 
reads 
\begin{eqnarray} \label{eq65}
B_3^{(ZWA)}=V_p^2+2{\tilde R}_pA_pV_p+2A_p^3\,.
\end{eqnarray}

\begin{figure}[t!]
\begin{center}
\includegraphics[width=8cm, clip=]{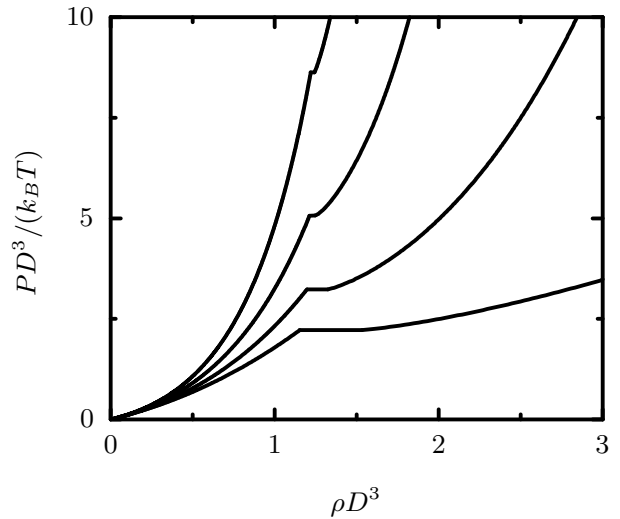}
\caption{Equation of state of a monodisperse fluid consisting of rectangular Zwanzig  
platelets of surface size $D \times D$ and thickness $L$ (see Fig.~\ref{fig14} and 
Eqs.~(\ref{eq53}) - (\ref{eq64})) for various aspect ratios \cite{harn:02c}: 
$L/D=0.01, 0.1, 0.2, 0.3$ (from bottom to top). The horizontal lines are tie-lines 
illustrating isotopic-nematic two-phase coexistence.}
\label{fig15}
\end{center} 
\end{figure}

As a direct application of the fundamental measure theory one may consider the 
possibility of an isotopic to nematic phase transition in a monodisperse fluid of 
platelets, as a function of the aspect ratio $L/D$. The broken symmetry 
in the nematic phase may be characterized by the nematic order parameter 
[Eq.~(\ref{eq46})]:
\begin{eqnarray} \label{eq66}
S=\frac{1}{2\rho V}\int d^3 r\,\left(
2\rho_z({\bf r})-\rho_x({\bf r})-\rho_y({\bf r})\right)\,,
\end{eqnarray}
where $\rho=V^{-1} \int d^3 r\,(\rho_x({\bf r})+\rho_y({\bf r})+\rho_z({\bf r}))$.
In the nematic phase $S=1$ and $\rho_x=\rho_y$. The one-component version of the 
grand potential [Eq.~(\ref{eq53})] has been minimized with respect to the order 
parameter $S$ for each density $\rho$ \cite{harn:02c}. The resulting equations 
of state in the isotropic ($S=0$) and nematic ($S\neq 0$) phases are plotted 
versus $\rho$ in Fig.~\ref{fig15}, together with the horizontal 
tie-lines connecting the two phases, for several aspect ratios $L/D$. The
value of the effective packing fractions $\phi_I=\rho_I D^3$ and 
$\phi_N=\rho_N D^3$ of the coexisting phases are seen to be rather insensitive 
to the aspect ratio $L/D$, while the first-order phase transition narrows as 
$L/D$ increases, i.e., $\triangle \phi=\phi_N-\phi_I$ decreases with increasing 
$L/D$. The reduced pressure $PD^3/(k_BT)$, however, increases rapidly 
with $L/D$, as one might expect.
%
%
%
%

\begin{figure}[t!]
\begin{center}
\includegraphics[width=8cm, clip=]{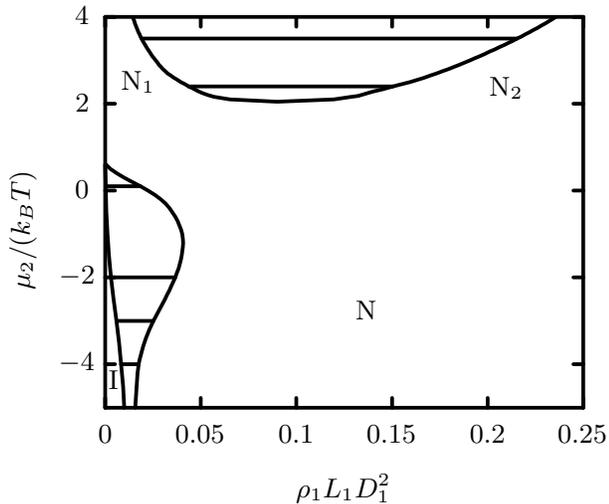}
\caption{Phase diagram of a fluid consisting of a binary mixture of large thin 
Zwanzig platelets [$L_1/D_1=0.01$] and small Zwanzig platelets [$D_1/D_2=2$, 
$L_2/D_2=0.16$] according to Eqs.~(\ref{eq53}) - (\ref{eq64})
as a function of the chemical potential of the small platelets $\mu_2$ 
and the density of the large platelets $\rho_1$ \cite{harn:02c}. The straight solid lines 
are tie-lines illustrating isotropic-nematic (I-N) and nematic-nematic (N$_1$-N$_2$) 
two-phase coexistence, respectively. The chemical potential at I-N coexistence of a 
monodisperse fluid consisting of the small platelets is $\mu_2/(k_BT)=0.68$.}
\label{fig16}
\end{center} 
\end{figure}

One may next consider the richer case of highly asymmetric binary mixtures of
large and small platelets. Polydispersity, both in diameters and in thickness,
is intrinsic to the aforementioned experimental gibbsite samples. In the case of 
mixtures of spherical colloidal particles large size asymmetry may lead to 
depletion-induced phase separation. In the case of platelets depletion-induced 
segregation competes with nematic ordering \cite{rowa:02}, as is also the case for 
mixtures of long and short rods, or of thin and thick rods. Figure \ref{fig16} 
displays a phase diagram of binary mixtures of square platelets constructed as function
of the chemical potential $\mu_2$ of the small platelets and the number density 
$\rho_1$ of the large platelets ($D_1>D_2$). The tie-lines are horizontal because 
of equality of $\mu_2$ of the coexisting phases. For large negative chemical potential 
$\mu_2$, the systems exhibit a first-oder isotropic (I) to nematic (N) phase transition 
and the density gap at the isotropic to nematic phase transition broadens with increasing 
$\mu_2$ in agreement with calculations for binary mixtures of infinitely thin platelets 
\cite{harn:02d}. At high densities in the nematic phase, the small platelets are nearly 
excluded and hence the large platelets are in equilibrium with a reservoir of small 
platelets. This implies that the pressure of the nematic phase is equal to that of a 
reservoir of small platelets at a chemical potential $\mu_2$. The isotropic-nematic 
two-phase coexistence regime is bounded by an upper critical point above which a single  
stable nematic phase is found. The nematic phase demixes into two nematic phases 
(N$_1$, N$_2$) at sufficiently large values of $\mu_2$. 

Upon increasing the aspect ratio $L_2/D_2$ of the small platelets at fixed $L_1$ and 
$D_1$, the lower critical point of the nematic-nematic two-phase coexistence region 
shifts to smaller values of $\mu_2$, until the nematic-nematic and isotropic-nematic 
two-phase coexistence regions start to overlap, giving rise to a triple point at which 
two nematic phases (N$_1$, N$_2$) coexist with an isotropic phase \cite{harn:02c}. 
%
%
%
%

\begin{figure}[t!]
\begin{center}
\includegraphics[width=8.5cm, clip=]{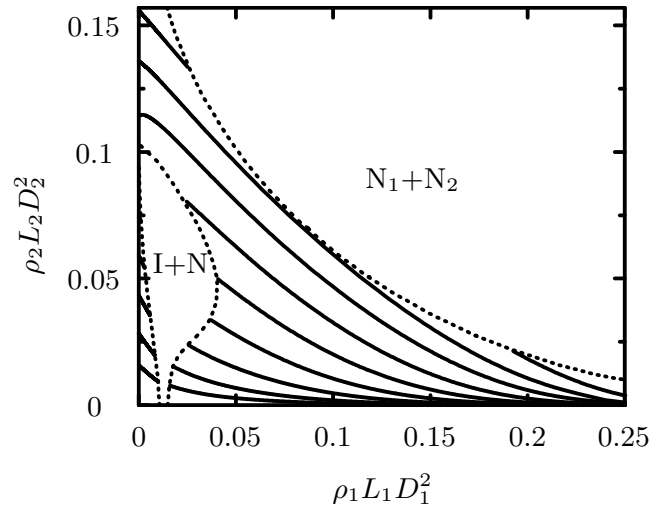}
\caption{Phase diagram of a binary platelet fluid, corresponding to Fig.~\ref{fig16}, 
in the density-density, i.e., $\rho_1$-$\rho_2$  plane \cite{harn:02c}. The dotted lines 
indicate phase boundaries. The chemical potential of the small thick platelets is kept 
fixed for each solid line and increases from bottom to top: 
$\mu_2/(k_BT)=-5, -4, -3, -2, -1, 0, 1, 1.9, 3$.}
\label{fig17}
\end{center} 
\end{figure}

Figure \ref{fig17} displays an alternative representation of the phase diagram shown 
in Fig.~\ref{fig16} in terms of the number densities of the large and the small platelets 
$\rho_1$ and $\rho_2$, respectively. The figure illustrates how the composition of the 
mixture varies upon increasing the chemical potential (and hence the reservoir density) 
of the small platelets. Whereas the density of the thin platelets $\rho_1$ is always 
smaller in the isotropic phase than in the nematic phase along the coexistence curve, 
a density inversion for the thick platelets has been found. The density of the thick 
platelets $\rho_2$ is larger in the isotropic phase than in the coexisting nematic 
phase for small chemical potentials $\mu_2$, while $\rho_2$ is smaller in the isotropic 
phase than in the coexisting nematic phase for large chemical potentials $\mu_2$.
This remarkable phenomenon of isotropic-nematic density 
inversion has been observed experimentally for a dispersion of sterically stabilized 
gibbsite platelets \cite{kooi:01a}. The calculated density profile $\rho_2(z)$ of 
thick platelets at the isotropic-nematic interface exhibits oscillations \cite{bier:04}.
The interfacial tension for platelets with their mean bulk orientation parallel to the 
interface is smaller than for the corresponding perpendicular configuration. The 
calculated interfacial tension \mbox{$\gamma\approx 10$ nN/m} for gibbsite platelets 
agrees with the measured value \cite{kooi:01,beek:06a}. Both the large platelet and 
small platelet density profiles exhibit pronounced oscillations on one side of the 
nematic-nematic interface, provided the chemical potentials are sufficiently 
high. Upon approaching the lower critical point of the nematic-nematic demixing 
region (see Fig.~\ref{fig16}) the oscillations vanish at a Fisher-Widom line and the 
interface broadens \cite{bier:04}.

%
%
%
%

A second-order virial approximation [Eq.~(\ref{eq55}), without the last term] has been 
used to investigate the phase diagram of binary mixtures of rods and  platelets
\cite{roij:94,soko:97}. It has been found that mixing rods  and platelets can stabilize 
a biaxial nematic phase in which the symmetry axes of particles of different species point 
along mutually perpendicular directions. This phase behavior was later confirmed in 
calculations for a polydisperse mixture of rods and platelets using fundamental measure 
theory \cite{rato:02}. A virial approximation has also been used to study wetting 
and capillary nematization of platelet and rod fluids in contact with planar and 
geometrically structured substrates \cite{harn:02d,harn:04}. The predicted complete 
wetting of the substrate-isotropic liquid interface by a uniaxial nematic platelet film 
has been found later in a colloidal dispersion of sterically stabilized gibbsite platelets
\cite{beek:06a}. The wetting behavior of the nematic gibbsite phase has been revealed 
by polarized light microscopy.

The relaxation dynamics of Zwanzig platelets has been investigated by means of 
dynamic density functional theory \cite{bier:07}. The driving force for time 
dependence has been expressed in terms of gradients of the local chemical potential. The 
formalism has been applied to model an initially homogeneous stable or metastable isotropic 
fluid which is perturbed by switching a two-dimensional array of Gaussian laser beams.
Switching on the laser beams leads to an accumulation of platelets in the beam centers.
If the initial chemical potential and the laser power are large enough, a preferred 
orientation of the particles occurs, breaking the symmetry of the laser potential.
After switching off the laser beams again, the system can follow different relaxation 
paths. It either relaxes back to the homogeneous isotropic state or it forms an 
approximately elliptical high-density core which is elongated perpendicular to
the dominating orientation in order to minimize the surface free energy. For large 
supersaturations of the initial isotropic fluid, the high-density cores of 
neighboring laser beams of the two-dimensional array merge into complex superstructures.

Recently, nonequilibrium steady states in an open system connecting two reservoirs of 
platelike colloidal particles have been investigated using a dynamic density functional 
theory for the Zwanzig model \cite{bier:08}. Inhomogeneities of the local chemical 
potential generate a diffusion current which relaxes to a nonvanishing value if the 
two reservoirs coupled to the system sustain different chemical potentials. The relaxation 
process of initial states towards the steady state turned out to comprise two regimes: 
a smoothening of initial steplike structures followed by an ultimate relaxation of the 
slowest diffusive mode. The position of a nonequilibrium interface and the particle 
current of steady states depends nontrivially on the structure of the reservoirs due 
to the coupling between translational and orientational degrees of freedom of the fluid.

\subsection{Equilibrium properties of charged Zwanzig platelets}
Bulk properties and free interfaces of mixtures of charged platelets and salt have been 
studied within the density functional theory for the Zwanzig model \cite{bier:05}.
The excess free energy functional has been decomposed into the fundamental measure 
contribution [Eq.~(\ref{eq58})] and an electrostatic part which has been derived by 
functional integration of an extension of the Debye-H\"uckel pair distribution 
function with respect to the interaction potential. The calculations have shown 
that the isotropic and nematic binodals are shifted to higher platelet densities 
upon increasing the platelet charge. The Donnan potential between the coexisting 
isotropic and nematic phases has been inferred from bulk structure calculations. 
Nonmonotonic density and nematic order parameter profiles have been found at a 
free interface interpolating between the coexisting isotropic and nematic phases.
Moreover, electrically charged layers form at the free interface leading to monotonically 
varying electrostatic potential profiles. For fixed salt density, the interfacial 
tension decreases upon increasing the macroion charge.

Analytically and numerically calculated surface phase diagrams of charged Zwanzig 
platelets and salt in contact with a charged planar substrate exhibit first-order 
wetting for sufficiently small platelet charges and isotropic bulk order as well as 
first-order drying for sufficiently large platelet charges and nematic bulk order 
\cite{bier:06}. A crossover from monotonic to nonmonotonic electrostatic potential 
profiles upon varying the surface charge density has been observed. Nonmonotonic 
electrostatic potential profiles are equivalent to the occurrence of charge inversion.
Due to the presence of both the Coulomb interactions and the hard-core repulsions, 
the surface potential and the surface charge do not vanish simultaneously, i.e., the point 
of zero charge and the isoelectric point of the surface do not coincide.

\section{Spatially inhomogeneous bulk phases} 
The Zwanzig model offers the advantage that the difficult determination of spatially
inhomogeneous bulk phases becomes numerically feasible \cite{rato:03,bier:04,rato:04,rato:07}. 
The predicted phase behavior is in qualitative agreement with the results obtained for
freely rotating parallelepipeds using computer simulations \cite{john:05}. In order to 
find spatially inhomogeneous bulk phases such as columnar and smectic phases, the spatially 
homogeneous solutions of the Euler-Lagrange equations for the fundamental measure theory 
of the Zwanzig model [Eqs.~(\ref{eq53}) - (\ref{eq64})] are perturbed by adding a narrow 
Lorentz peak and iterated using a Picard scheme. If the latter converges towards the spatially 
homogeneous solution (isotropic or nematic phase), one concludes that this is indeed the 
equilibrium state because it is stable under density variations. This method is the 
numerical analog of a bifurcation analysis since a narrow Lorentz peak has a broad Fourier 
spectrum. In the case of the phase diagram for the binary mixtures of large thin and small 
thick platelets shown in Figs.~\ref{fig16} and \ref{fig17} it has been found that the nematic 
phase N$_1$ undergoes a phase transition into a columnar phase, where parallel columns 
of small platelets are surrounded by single platelets with orientations perpendicular 
to the column axis \cite{bier:04}. For small values of $\rho_1$, the nematic to columnar 
phase transition is weakly first-order. The nematic-columnar boundary line changes to 
strongly first-order at the tricritical point where the nematic-nematic boundary line 
ends.

\begin{figure}[t!]
\begin{center}
\includegraphics[width=8cm, clip=]{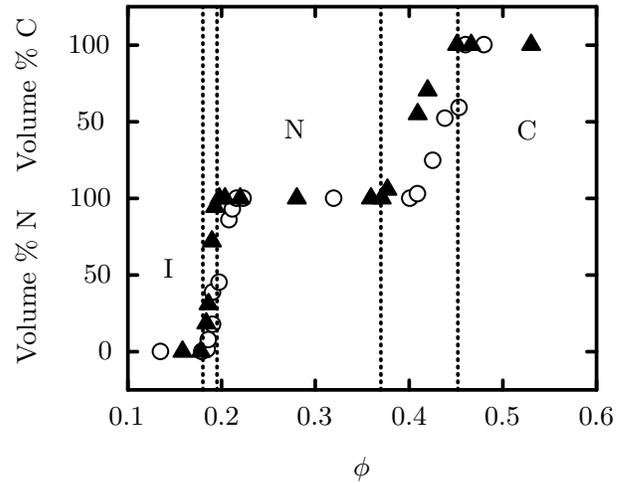}
\caption{Phase diagram of two gibbsite suspensions which differ by the degree 
of polydispersity of the platelet radius (triangles, \mbox{17 \%} and circles, 
\mbox{25 \%}) \cite{kooi:00d}. The relative volume fraction of the nematic (N) 
and columnar (C) phase are plotted as function of the platelet volume fraction $\phi$.
For low volume fraction the isotropic phase (I) is stable. The dotted lines mark the 
boundaries of the two-phase coexistence regions of the system with \mbox{17 \%} 
polydispersity of the platelet radius.}
\label{fig18}
\end{center} 
\end{figure}

Over the past few years, spatially homogeneous and inhomogeneous bulk phases of 
the aforementioned suspensions of colloidal gibbsite platelets have been investigated. 
For example, van der Kooij et al. \cite{kooi:00d} have studied two systems which differ 
by the degree of polydispersity of the radius of the platelets (\mbox{17 \%} and 
\mbox{25 \%}, respectively).
The phase behavior of the platelet suspensions as function of the volume fraction $\phi$
of the platelets is shown in Fig.~\ref{fig18}. Isotropic-nematic two-phase coexistence
has been observed just below $\phi=0.2$, yielding an isotropic upper phase and a 
birefringent nematic bottom phase in the tubes containing the suspensions. Macroscopic 
phase separation has been found to be complete within 12 hours. The width of the 
biphasic gap  $\triangle \phi=\phi_N-\phi_I$ increases with increasing the degree 
of polydispersity of the platelet radius. Upon increasing the volume fraction of the 
platelets to  $\phi \approx 0.4$, both suspensions enter a biphasic region where a
nematic phase coexists with a more concentrated birefringent columnar phase. A 
comparison of the experimentally observed volume fractions at the isotropic to nematic 
and nematic to columnar phase transitions with computer simulation data for mondisperse 
platelets \cite{veer:92} has shown that the transitions in the experiment are shifted to
lower volume fractions which might be due the polydispersity in size and shape of the 
gibbsite platelets \cite{kooi:00d}. The stability of the observed columnar phase is 
remarkable in view of predictions for the terminal size polydispersity of about \mbox{10 \%}
for the crystal phase of hard spheres and the terminal length polydispersity of 
about \mbox{18 \%} for the smectic phase of hard rods. Small-angle x-ray scattering 
has revealed a hexagonal intercolumnar ordering in the columnar phase \cite{petu:05}.
The system forms a powder consisting of true long-range-ordered columnar crystallites, 
where the relatively large free space between the columns allows for the accommodation
of rather highly polydisperse platelets. However, under the influence of gravitational 
compression very little space is available. The geometrical frustration induced by the 
platelet polydispersity can suppress the ordering upon increasing density and favor 
hexaticlike structuring. The resulting novel columnar phase has been found at the 
bottom of a tube containing a gibbsite suspension \cite{petu:05}. The hexatic phase 
is characterized by short-range translational order while its bond-orientational 
order is long ranged. 

Moreover, polydisperse colloidal gibbsite platelets have been
reported to form an opal-like columnar crystal with iridescent Bragg reflections 
\cite{beek:07}. It has been shown that the formation process of the iridescent 
phase under slow sedimentation can be accelerated by orders of magnitude using a 
centrifugation without arresting the system in a disordered glassy state. It has 
been argued that the formation of the hexatic phase with a finite position correlation 
length may play a role in the self-organization of the platelets \cite{beek:07}.

%
%
%
%

\section{Summary and outlook}
This work has been devoted to the investigation of the structure and 
thermodynamics of platelet dispersions. The shape of the particles and the 
polydispersity in size can be revealed 
by cryogenic transmission electron microscopy [Fig.~\ref{fig1}] and the 
form factor [Eqs.~(\ref{eq1}) - (\ref{eq4})] measured by light scattering, 
small-angle neutron scattering, or small-angle x-ray scattering [Figs.~\ref{fig3} 
and \ref{fig4}]. The combination of these methods has demonstrated that 
well-defined and perfectly dispersed single lamellar nanoplatelets of the 
cheapest polyolefin, namely polyethylene, can be made in water \cite{webe:07}. 
Due to their unusual size, structure, and thermal history, these platelets 
will contribute to fundamental and long-standing issues of polymer crystallization 
\cite{chen:07}.

The interaction site integral equation theory [Eqs.~(\ref{eq5}) - (\ref{eq13})] 
is quite successful in describing experimental structure factors of fluids 
containing uncharged platelets [Fig.~\ref{fig5}] or uniformly charged platelets 
[Fig.~\ref{fig6}]. Rigid dendrimers [Fig.~\ref{fig2}] may serve as model systems for 
interacting monodisperse particles in statistical physics \cite{harn:07}, while 
the colloidal platelets synthesized so far are highly polydisperse in size.

The inverse structure factor extrapolated to vanishing scattering vectors as 
predicted by both scaled particle theory and fundamental measure theory is 
systematically smaller than the experimental data for platelike stilbenoid dendrimers 
of the third generation and the prediction of the interaction site integral 
equation theory [Fig.~\ref{fig9} and Eqs.~(\ref{eq18}) - (\ref{eq24})]. The substantial 
differences observed between the experimental data and the results of scaled particle 
theory and fundamental measure theory confirm earlier caveats concerning the 
applicability of these theories to freely rotating nonspherical particles 
\cite{over:05a} and are due to the fact that both theories do not yield the correct 
third virial coefficient. Hence there is a clear need to improve both scaled particle 
theory and fundamental measure theory for platelets.

The different charge densities on the rim and the face of laponite platelets may 
lead the significant difference between the effective pair potentials and structure 
factors of laponite and polyelectrolytes [Figs.~\ref{fig7} and \ref{fig8}]. However, 
at the present stage there is no definite answer to this issue because no theoretical 
calculations are available yet. An integral equation approach based on the reference 
interaction site model with different charges on the face and the rim of the platelets 
is expected to be helpful for the understanding of the microscopic structure of 
laponite suspensions. An important point that is common to many experimental studies 
of laponite suspensions is the analysis of obvious nonequilibrium states reached 
after passing the gel line. As a matter of fact, considerable efforts have been 
devoted to the investigation of the exact position of the gel line. The experimental 
difficulties of the investigation of such time-dependent states are at hand. Moreover, 
no clear information on the nature of the interparticle interaction potential can 
be derived from essentially nonequilibrium states. 

Density functional theory [Eqs.~(\ref{eq26}) - (\ref{eq34})] has been used to study the 
phase behavior of thin hard platelets [Fig.~\ref{fig10}] in three dimensions. The system 
exhibits a first-order isotropic to nematic phase transition with a small density 
jump at two-phase coexistence and a rather low nematic order parameter in the 
nematic phase. Within a geometry-based density functional theory the excess free 
energy is expressed in terms of weighted densities [Eqs.~(\ref{eq35}) - (\ref{eq44})].
However, this fundamental measure theory does not yield the correct third virial 
coefficient due to the occurrence of lost cases. These lost cases can be compensated 
by overcounting the cases with triple intersection such that the resulting equation 
of state in both the isotropic and the nematic phase as well as the location of the 
isotropic to nematic phase transition agree with computer simulation data [Fig.~\ref{fig11}]. 

Density functional theory has also been used to study binary colloidal fluids consisting 
of hard spheres and thin platelets in their bulk and near a planar hard substrate. 
This system exhibits liquid-liquid coexisting of a phase that is rich in spheres 
(poor in platelets) and a phase that is poor in spheres (rich in spheres) 
[Fig~\ref{fig13}]. An interesting issue that has not been addressed theoretically 
is how the roughness of platelets modifies depletion interactions \cite{zhao:07}.

Although the particle shape is equivalent, overlapping pair 
configurations are different for freely rotating platelets in three dimension as
compared to the strictly two-dimensional case. Fundamental measure theory yields 
only an approximate expression for the Mayer function of strictly two-dimensional
hard platelets [Fig.~\ref{fig12}]. The understanding of the deconvolution of the 
Mayer function for platelets in two dimensions needs to be improved \cite{capi:07}.

Zwanzig's model of parallelepipeds with only three allowed orientations [Fig.~\ref{fig14}] 
has been generalized to the case of monodisperse, bidisperse, and polydisperse 
systems of platelets. In view of the uncertainties concerning the convergence of 
the virial series [Eqs.~(\ref{eq53}) - (\ref{eq57})], a fundamental measure theory 
has been used to derive a free energy functional [Eqs.~(\ref{eq58}) - (\ref{eq64})].
This fundamental measure theory for the Zwanzig model yields the exact second and 
third virial coefficients irrespective of the shape and size polydispersity of the 
particles. In the monodisperse case, the isotropic to nematic phase transition 
narrows with increasing aspect ratio of the platelets [Fig.~\ref{fig15}]. The bulk 
phase diagram for a representative example of binary hard platelet fluids
[Figs.~\ref{fig16} and \ref{fig17}] involve an isotropic and one or two nematic 
phases of different concentrations as well as a columnar phase. 

The Zwanzig model offers the advantage that the difficult determination of spatially 
inhomogeneous bulk phases and density profiles becomes numerically feasible. The calculated 
low interfacial tensions are in agreement with experimental data for gibbsite platelets. 
Moreover, the predicted wetting behavior of platelet fluids in contact with substrates 
has been confirmed experimentally \cite{beek:06a}. It is appealing to base the 
construction of theories for the dynamics of platelets on the equilibrium fundamental 
measure functional for the Zwanzig model \cite{bier:07}. In addition to the technological 
relevance of this issue (e.g., sediment transport processes in hydraulic engineering), 
the investigation of dynamic properties of platelet fluids provides the opportunity 
to study the coupling of translational and rotational motion as well as the time 
dependence of the orientation of particles induced by external fields. 

The liquid-crystalline phase behavior of colloidal gibbsite platelets 
has been studied over the past few years. It has been observed that suspensions
of both sterically and charge-stabilized gibbsite platelets at sufficiently 
high concentrations display iridescence. From small-angle x-ray scattering 
experiments it follows that these iridescent phases have a columnar structure 
[Fig.~\ref{fig18}] with hexagonal intercolumnar ordering. This is quite remarkable
because the gibbsite platelets have a polydispersity of \mbox{19 \%} in diameter
and \mbox{27 \%} in thickness. In general, size polydispersity of colloidal
particles is expected to suppress the formation of ordered phases. 

For the case of monodisperse hard platelets an accurate description of a columnar 
liquid crystal phase at high packing fractions has been presented using an improved 
free volume theory \cite{wens:04a,lai:06}. It has been shown that orientational 
entropy of the platelets in the one-dimensional columns leads to a different 
high-density pressure compared to the prediction from traditional cell theory. 
Quantitative agreement has been found with Monte Carlo simulation results for 
various thermodynamic properties. In view of this progress
made in developing an accurate descriptions of a columnar liquid crystal 
consisting of monodisperse hard platelets, future work may focus 
on the understanding of the influence of platelet size polydispersity, gravity, 
and external fields on the phase behavior. Discotic liquid 
crystal science is still a rather young research field compared with the more 
developed field of liquid crystals consisting of rodlike molecules. Nevertheless,
many applications for discotic liquid systems in areas such as photovoltaic
and optical compensation films for displays have been investigated over the last 
years. However, the microscopic understanding of the properties of columnar phases 
and the alignment at surfaces needs to be improved in particular in view of the 
emerging research area of smart low-dimensional building blocks consisting of 
discotic liquid crystals.

\end{document}